\documentclass[manuscript,screen,acmsmall]{acmart}

\AtBeginDocument{%
  \providecommand\BibTeX{{%
    \normalfont B\kern-0.5em{\scshape i\kern-0.25em b}\kern-0.8em\TeX}}}

\setcopyright{acmlicensed}
\copyrightyear{2025}
\acmYear{2025}
\acmDOI{XXXXXXX.XXXXXXX}

\usepackage{array}
\newcolumntype{S}{>{\arraybackslash}m{3cm}}
\newcolumntype{L}{>{\arraybackslash}m{5.5cm}}
\usepackage{multirow}
\usepackage{ltablex}

\begin{document}

\title[To Recommend or Not to Recommend]{To Recommend or Not to Recommend: Designing and Evaluating AI-Enabled Decision Support for Time-Critical Medical Events}

\author{Angela Mastrianni}
\orcid{0000-0001-8179-0101}
\affiliation{%
  \institution{Drexel University}
  \city{Philadelphia}
  \state{PA}
  \country{USA}
}
\email{amastrianni12@gmail.com}

\author{Mary Suhyun Kim}
\email{MSKIM@childrensnational.org}
\orcid{0009-0006-7025-6636}
\affiliation{%
  \institution{Children's National Hospital}
  \city{Washington}
  \state{D.C.}
  \country{USA}
}

\author{Travis M. Sullivan}
\orcid{0000-0002-4399-7037}
\affiliation{%
  \institution{Children's National Hospital}
  \city{Washington}
  \state{D.C.}
  \country{USA}
}

\author{Genevieve Jayne Sippel}
\orcid{0000-0002-7952-766X}
\affiliation{%
  \institution{Children's National Hospital}
  \city{Washington}
  \state{D.C.}
  \country{USA}
}

\author{Randall S. Burd}
\email{RBurd@childrensnational.org}
\orcid{0003-4465-9117}
\affiliation{%
  \institution{Children's National Hospital}
  \city{Washington}
  \state{D.C.}
  \country{USA}
}

\author{Krzysztof Z. Gajos}
\orcid{0000-0002-1897-9048}
\affiliation{%
  \institution{Harvard University}
  \city{Allston}
  \state{Massachusetts}
  \country{USA}}
\email{kgajos@seas.harvard.edu}

\author{Aleksandra Sarcevic}
\orcid{0000-0003-2126-8527}
\affiliation{%
  \institution{Drexel University}
  \city{Philadelphia}
  \state{PA}
    \country{USA}
}
\email{aleksarc@drexel.edu}

\renewcommand{\shortauthors}{Mastrianni, et al.}

\begin{abstract}
AI-enabled decision-support systems aim to help medical providers rapidly make decisions with limited information during medical emergencies. A critical challenge in developing these systems is supporting providers in interpreting the system output to make optimal treatment decisions. In this study, we designed and evaluated an AI-enabled decision-support system to aid providers in treating patients with traumatic injuries. We first conducted user research with physicians to identify and design information types and AI outputs for a decision-support display. We then conducted an online experiment with 35 medical providers from six health systems to evaluate two human-AI interaction strategies: (1) AI information synthesis and (2) AI information and recommendations. We found that providers were more likely to make correct decisions when AI information and recommendations were provided compared to receiving no AI support. We also identified two socio-technical barriers to providing AI recommendations during time-critical medical events: (1) an accuracy-time trade-off in providing recommendations and (2) polarizing perceptions of recommendations between providers. We discuss three implications for developing AI-enabled decision support used in time-critical events, contributing to the limited research on human-AI interaction in this context.
\end{abstract}


\maketitle

\section{Introduction}
Computational methods can diagnose medical diseases with high levels of accuracy in medical research, especially when using artificial intelligence (AI) \cite{fatima2017survey}. A critical challenge in using these methods to improve diagnostic and decision-making processes is supporting medical providers in interpreting the outputs of AI-enabled systems \cite{van2023measurements}. Because providers are responsible for determining treatment plans, their interpretation of system output is more important than the output itself \cite{vasey2021association}. Many AI-enabled decision-support systems provide recommendations, adding explanations to support users in interpreting the recommendations \cite{gunning2019xai}. However, recent research in human-AI interaction has proposed a paradigm shift from recommendation-driven support to information-driven support, which provides users with information synthesis to facilitate hypothesis testing \cite{miller2023explainable, gajos2022people, zhang2024rethinking, zhang2024beyond}. Information-driven support could help providers evaluate potential diagnoses, while avoiding issues commonly found with recommendations, such as over-reliance and reduced agency \cite{miller2023explainable, zhang2024rethinking, zhang2024beyond}. Although initial empirical studies suggest that this approach can improve decision making, knowledge about the effectiveness of information-driven support, especially during instances of time-critical clinical decision making, remains limited. In addition, the role of recommendations within the information-driven support paradigm is still an open research area. Some information-driven strategies provide only information and not recommendations \cite{gajos2022people}. Others provide recommendations toward the end of the decision-making process to confirm or challenge the initial ideas of users \cite{zhang2024beyond}.

To examine the effectiveness of information-driven support for time-critical decision making and the role of recommendations, we designed and evaluated a decision-support system within the context of pediatric trauma resuscitation. During resuscitations, providers treat critically injured patients in the emergency department, assessing their need for life-saving interventions (LSIs). We focused on developing a system to support decisions about two LSIs: blood transfusion and neurosurgical intervention. These LSIs are administered when treating two of the most frequent causes of preventable deaths in pediatric trauma resuscitation, severe blood loss and traumatic brain injury \cite{drake2020establishing}. We started our study by surveying 19 physicians from emergency medicine and surgery to understand the types of information that could support decision making during resuscitations. Most physicians thought the decision-support system should communicate the recommended interventions, the patient’s probability of receiving an LSI, and the patient information used by the decision-support model. From these results, we focused on comparing two human-AI interaction strategies: AI information synthesis and AI information and recommendations. Based on interviews with nine physicians and prior literature, we designed two displays for the system: (1) a display with AI information synthesis and (2) a display with AI information and recommendations.

To compare the effectiveness of the two human-AI interaction strategies on time-critical decision making, we conducted an online experiment with 35 surgeons, physicians, and advanced practice providers from six healthcare organizations. In the experiment, participants had a limited amount of time to make decisions about the need for LSIs in 12 simulated patient vignettes. Participants were exposed to three types of AI support: (1) no AI support, (2) AI information synthesis, and (3) AI information and recommendations. Because our focus was on evaluating the impacts of different human-AI interaction strategies on decision making rather than on assessing the accuracy of any specific computational model, we simulated an AI system to generate and control the recommendations and information synthesis. We had two research hypotheses:
\begin{itemize}
    \item H1: Providing AI information synthesis will improve providers' decision making (compared to no AI support).
    \item H2: Providing AI information and recommendations will improve providers' decision making (compared to no AI support).
\end{itemize}
We found evidence supporting our second hypothesis that providing AI information and recommendations improves decision making. Participants were more likely to make correct decisions when provided AI information and recommendations (compared to receiving no AI support), even though the recommendations were occasionally incorrect. We did not find evidence for the first hypothesis that providing AI information synthesis alone significantly improved decision making compared to no AI support. We also did not find a significant difference in decision making between AI information and AI information and recommendation. Findings from the experiment suggest that both strategies may have functioned as a type of process-oriented support \cite{zhang2024beyond}, with patient information synthesized on the display as it was being presented in the vignette. Participants often made initial decisions as information was being presented, with some participants highlighting that they used the information synthesis to identify abnormalities and check their logic. Because recommendations appeared at the end of the vignette, they often served as a cognitive forcing function \cite{buccinca2021trust}. Receiving either type of AI support did not increase the time taken to make decisions. Although AI information and recommendations was the most preferred type of AI support, participants had polarizing perceptions and uses of the recommendations. Some providers appreciated the ability to use the recommendations to check their decisions, while others thought the recommendations lacked nuance and could bias their decision making. 

From our work, we make three contributions to CSCW:
\begin{itemize}
    \item We show that providing AI information and recommendations can improve decisions about life-saving interventions, even if the recommendations are occasionally incorrect. These results add new knowledge to the limited evidence that human-AI collaboration can enhance decision making of domain experts in time- and resource-constrained settings.
    \item We highlight two socio-technical barriers to implementing AI recommendations in time-critical medical events: (1) determining when to present AI recommendations and (2) supporting multiple human-centric objectives in addition to improving decision making. Controlling the points when recommendations are provided in the clinical workflow during time-critical medical events is more challenging due to the dynamic and uncertain nature of patient information. Although recommendations may improve decision making, recommendations may also negatively impact provider collaboration and agency.
    \item We discuss three implications for developing AI-enabled decision support used during time-critical events: (1) support human decision makers in critically evaluating AI-generated information synthesis, (2) consider other information-based strategies, beyond only recommendations, to support decision makers in managing accuracy-time trade-offs, and (3) develop clear policies around the responsibility and liability of decision makers when adopting AI-enabled tools. 
\end{itemize}

\section{Related Work}
We discuss related work on clinical decision-support systems and strategies for designing human-AI interaction more broadly.

\subsection{Clinical Decision-Support Systems}
Despite their promise, clinical decision-support systems have faced issues with adoption and effectiveness \cite{sittig2008grand}. Effective clinical decision-support systems can be challenging to develop, especially for complex social-technical environments \cite{jacobs2021designing, yang2016investigating, yang2019unremarkable, yang2023harnessing, kaltenhauser2020you, zhang2022get, kluber2020experience, ghosh2023framing}. One barrier to adoption is the negative perceptions of decision-support systems held by medical providers. Providers may believe that these systems are not needed or do not provide adequate support in patient cases with missing or ambiguous information \cite{yang2016investigating}. They may also have concerns that these systems will require additional cognitive effort in already time- and resource-constrained scenarios \cite{jacobs2021designing, kaltenhauser2020you, zhang2022get, bach2023if}. For example, ophthalmologists had negative perceptions of bias mitigation strategies incorporated within a decision-support system because the strategies added extra steps to their workflows (e.g., by asking them to justify certain decisions) \cite{bach2023if}. Finally, providers may have concerns that these systems will infringe on their agency \cite{valenta2010physician, wang2021brilliant, berge2023designing}.

Clinical decision-support systems can be knowledge-based or non-knowledge-based, depending on how the system generates decision support \cite{sutton2020overview}. Most early clinical decision support systems were knowledge-based, relying on literature-based and local practice-based evidence to develop decision-support guidelines \cite{sim2001clinical}. Advancements in AI have increased the development of non-knowledge-based systems that infer patterns from large quantities of low-level data \cite{yu2018artificial}. Because decision support is no longer directly derived from evidence-based rules, supporting providers in interpreting and validating output from the ``black-box" models has become a challenge \cite{sutton2020overview}. When providers cannot accurately assess the quality of the output, they may ignore correct advice from the system or follow incorrect advice. For example, radiologists in one study found it challenging to determine the correctness of AI predictions, leading to poor decisions when inaccurate predictions were provided \cite{yu2024heterogeneity}. When using a decision-support system, users interact with both the information from the event (observed space) and information from the system (prediction space) \cite{rastogi2022deciding}. The user may be affected by different biases from their interactions with each space. For example, availability bias may affect perceptions of the event, while predictions from the system can lead to anchoring bias (e.g., insufficient deviation from inaccurate predictions).

Prior research studied strategies for supporting providers in interacting with AI-enabled clinical decision-support systems \cite{jacobs2021designing, burgess2023healthcare, yang2023harnessing, yang2023harnessing}. One strategy involves incorporating tools that allow providers to refine the output of the decision-support system \cite{cai2019human}. This strategy was explored in radiology, allowing radiologists to mark useful images retrieved by the system to obtain additional, similar results. Other strategies are inspired by evidence-based medicine \cite{jacobs2021designing, yang2023harnessing}. To help providers validate suggestions from AI models, these strategies involve showing information on model validation \cite{jacobs2021designing} and displaying evidence from biomedical literature that supports or refutes the suggestion \cite{yang2023harnessing}. Evidence-based medicine strategies have been explored with physicians from different specialties \cite{jacobs2021designing, yang2023harnessing}, with emergency medicine physicians expressing concerns about the amount of time available to deeply engage with biomedical literature \cite{yang2023harnessing}. While most clinical settings have time constraints, providers in medical emergencies operate under extreme time pressure. Strategies that work for settings like radiology may not be appropriate for emergency medical settings because providers have limited time to engage with system outputs or interact with the system. We contribute to this line of research by studying the design of human-AI interaction within systems used in dynamic and time-critical medical settings.

\subsection{Human-AI Interaction Paradigms in Decision-Support Systems}
Clinical decision-support systems have often used recommendations to support decision making. However, recent research in the broader field of AI-supported decision-making has proposed a paradigm shift from recommendation-driven support to information-driven support. 

\subsubsection{Recommendation-Driven Support}
In recommendation-driven support, decision-support systems provide recommendations to support users in solving a problem \cite{zhang2024beyond}. To help users interpret and validate the recommendations, explanations are often presented with recommendations \cite{gunning2019xai}. Explanations can increase the confidence of medical providers in an AI system \cite{sivaraman2023ignore, panigutti2022understanding}, even if the explanation is perceived to be of low quality \cite{panigutti2022understanding}. Some explainable AI techniques provide global explanations that facilitate an understanding of the entire system, while others offer local explanations that facilitate understanding of a specific system recommendation ~\cite{antoniadi2021current}. Although explainable AI techniques can mitigate common biases in decision making \cite{wang2019designing}, providing explanations with recommendations has shown mixed results \cite{bussone2015role, vasconcelos2023explanations, bansal2021does, buccinca2021trust, panigutti2022understanding, lee2023understanding}. In some studies, providing explanations with recommendations reduced over-reliance on incorrect predictions or recommendations \cite{lee2023understanding, vasconcelos2023explanations}. In others, over-reliance on system outputs increased with explanations~\cite{bansal2021does}. Several factors may be contributing to these mixed results. First, the interpretation and use of recommendations may be influenced by the type of explanation provided \cite{lee2023understanding}. Salient feature-based explanations show the values of the features weighted most heavily by a computational model, while counterfactual explanations highlight how the model output would change with different input values. Second, findings from empirical studies may be influenced by the type of task performed by the participant \cite{buccinca2020proxy, vasconcelos2023explanations}. Relying on proxy tasks (e.g., asking participants to predict the output of a system) or subjective measurements can result in misleading results \cite{buccinca2020proxy}. Explanations may also be more effective for difficult or high-reward tasks because the benefit of engaging with the explanation may outweigh the cost \cite{vasconcelos2023explanations}. 

Receiving an AI recommendation before or after making an initial decision can also influence the acceptance of recommendations. For example, some cognitive forcing strategies intentionally delay showing users the AI recommendations or require making a provisional decision before viewing the recommendation. This approach can reduce acceptance of AI recommendations regardless of their accuracy \cite{buccinca2021trust, fogliato2022goes, park2019slow}. Users in clinical and non-clinical settings perceived systems with these cognitive forcing strategies as less useful than other AI systems \cite{buccinca2021trust, fogliato2022goes}.

\subsubsection{Information-Driven Support} Recent work has proposed alternative human-AI interaction strategies, such as evaluative AI \cite{miller2023explainable} and process-oriented support \cite{zhang2024beyond}, that focus on providing information synthesis instead of recommendations. While recommendation-driven support aims to solve a problem for users, information-driven support focuses on mitigating challenges in decision making by helping users synthesize information and assess hypotheses \cite{miller2023explainable, zhang2024beyond, zhang2024rethinking}. These strategies could better align with the abductive reasoning process used during decision making and reduce over-reliance on AI by increasing user agency \cite{miller2023explainable}. Although further research is needed to assess the effectiveness of these strategies, initial empirical studies have suggested that they can support decision making \cite{gajos2022people, danry2023don}. For example, compared to no AI support, only providing AI explanations and omitting recommendations improved learning gains of crowdsourcing platform users asked to make nutrition-related decisions \cite{gajos2022people}. In contrast, providing recommendations after a user had made a decision led to no learning gains. These results indicate that providing only AI explanations led to deeper cognitive engagement with the AI system output compared to also showing decision recommendations either before or after participants made their initial decisions. In addition to supporting decision making, information-driven support may provide users with more agency and autonomy than recommendation-driven support \cite{zhang2024beyond, zhang2024rethinking}. In one study, some intensive care and emergency medicine physicians preferred receiving information synthesis from AI-enabled systems to help them generate and test hypotheses instead of risk prediction scores that could challenge their authority \cite{zhang2024rethinking}. This study, however, did not evaluate how providing information synthesis or risk predictions affected decision making. We still have a limited understanding of the effectiveness of information-driven strategies in supporting decision making of domain experts performing real-world tasks in different settings \cite{zhang2024beyond}. In this paper, we study the effectiveness of these strategies in supporting time-critical clinical decision making. Because providers must rapidly evaluate and interpret information sources during medical emergencies, providing effective information synthesis may support their decision making without creating issues typically associated with AI recommendations (e.g., reduced agency).

\subsubsection{Effects of Time Pressure and Experience Level on Human-AI Collaboration} 
Time pressure can influence decision-making processes and outcomes. When making decisions under time pressure, people may be more likely to use recognition of past experiences and heuristics instead of more analytical strategies, such as systematically evaluating different options \cite{croskerry2003importance, rieskamp1999people, klein1993recognition}. Heuristics can support efficient decision making with incomplete information, but failed heuristics may lead to cognitive biases and errors \cite{dale2015heuristics, croskerry2003importance}. Because of heightened time pressure, poor access to information, and increased number of decisions, emergency medicine providers often rely on heuristics when making decisions \cite{croskerry2003importance}. Thirty types of failed heuristics have been identified in emergency medicine settings \cite{croskerry2002achieving}. One example is confirmation bias, which occurs when providers look for evidence supporting a diagnosis instead of evidence refuting it. Cognitive forcing strategies, such as diagnostic timeouts, may be effective in preventing failed heuristics by prompting providers to analytically engage with their decisions \cite{croskerry2013cognitive, lambe2016dual}. However, because these strategies generally force decision makers to slow down, they may have trade-offs in time-constrained settings.

Recent studies have examined how people interact and make decisions with AI under time pressure  \cite{swaroop2024accuracy, cao2023time, rastogi2022deciding}. \citet{rastogi2022deciding} provided participants with AI-generated risk predictions at the start of a task, finding that participants were more likely to sufficiently shift away from incorrect predictions when given more time to complete the task. In contrast, \citet{cao2023time} provided participants with AI recommendations after they had made initial decisions. Participants were more likely to shift from their initial decision and accept AI recommendations when given more decision-making time. Both studies suggest that time pressure may have contributed to anchoring, with participants less likely to deviate from recommendations provided at the start of a task \cite{rastogi2022deciding} or from initial decisions when recommendations were provided later \cite{cao2023time}. Swaroop et al. \cite{swaroop2024accuracy} investigated accuracy-time trade-offs with and without time pressure for three different types of AI support: (1) AI recommendations before an initial decision, (2) AI recommendations after an initial decision, and (3) a mix of AI-before, AI-after, and no AI support. When under time pressure, AI-before support led to faster response times than the other support types with similar levels of accuracy. This finding suggests that providing AI recommendations before an initial decision may be better for accuracy-time trade-offs when decisions are made under time pressure. Participants in all three studies were not domain experts and completed tasks that included student performance prediction \cite{rastogi2022deciding}, spatial reasoning \cite{cao2023time}, count estimation \cite{cao2023time}, and medication selection \cite{swaroop2024accuracy}. Our work contributes to this line of research in two ways. First, we investigate how domain experts interact with AI support when making decisions under time pressure. Emergency medicine providers are trained and have expertise in making high-risk decisions under time pressure, but their decision making can still be subject to cognitive biases. Second, we examine how users interact with information-driven AI support strategies when given a limited amount of time to make decisions.

\section{Phase I: Designing AI-Enabled Decision-Support Displays} 
In the first study phase, we administered a survey and interviewed physicians at our primary research site to determine the design of AI-enabled decision-support displays for trauma resuscitation. This study phase was approved by the Institutional Review Board at our primary research site.

\subsection{Primary Research Site}
Our primary research site was Children's National Hospital (CNH), a Level 1 pediatric trauma center in the United States. CNH treats about 600 injured children annually in the designated resuscitation rooms in the emergency department. During resuscitations, a team of providers stabilizes and evaluates the patient before determining a treatment plan and transporting the patient to their next destination in the hospital. Before patient arrival, the team may receive information about the incoming patient from the emergency medical services (EMS) transporting the patient to the hospital. This information can include the patient's age, mechanism of injury, vital signs, and neurological status measured by Glasgow Coma Scale (GCS). After patient arrival, the team examines the patient by following the Advanced Trauma Life Support (ATLS) protocol \cite{subcommittee}. At CNH, the resuscitations are co-led by a surgical fellow or senior resident and an emergency medicine physician. In cases with more severe injuries, a surgical attending and physicians from other specialties (e.g., anesthesia, critical care) may join the leadership team. A junior surgical resident examines the patient, while bedside nurses measure vital signs, obtain intravenous (IV) access, and assist with bedside tasks. The ATLS protocol contains two phases, the primary and secondary surveys. The team first assesses the patient's airway, breathing, circulation, and mental status (primary survey) and then examines the patient from head to toe for additional injuries (secondary survey). Leaders make critical decisions throughout the resuscitation, including determining the need for LSIs, such as blood transfusion and surgical procedures. Errors in reasoning can result in wrong diagnoses, delayed or inappropriate treatments, and even preventable deaths \cite{clarke2000objective, teixeira2007preventable, o2013opportunities, drake2020establishing}. Clinical reasoning errors can be more prevalent than errors with technical skills \cite{o2013opportunities}.

\subsection{Survey: Identifying Types of AI-Enabled Decision-Support Information}
From a review of prior literature, we first generated a list of five types of information that may aid medical providers in making decisions when interacting with decision-support systems: (1) recommended interventions \cite{yang2016investigating}, (2) patient information used by a system computational model \cite{kaltenhauser2020you, xie2020chexplain, lee2021human}, (3) comparisons between the current patient and past patients \cite{xie2020chexplain,jacobs2021designing}, (4) the process used to train and validate a system's model \cite{jacobs2021designing}, and (5) details about a model's performance \cite{jacobs2021designing}.  We then conducted an online survey, asking participants to (1) select the types of information they would consider when receiving AI-enabled decision support about an LSI (specifically blood transfusion for severe blood loss) and (2) rank the types of information from most to least important. In addition to the five types of information identified from the literature, we also included the patient's probability of receiving an LSI as another potential type of information based on feedback from clinical collaborators.

We distributed the online survey through a call-for-participation email sent to 79 surgical and emergency medicine residents, fellows, and attending physicians at our primary research site. Nineteen participants completed the survey (24\% response rate). Participants included two surgical attending physicians, 16 emergency medicine attending physicians, and one emergency medicine fellow (Appendix \ref{appendix_demopgrahics}). Participants had an average of 14 years of experience. Most participants thought that the AI-enabled decision-support system should communicate the (a) recommended interventions (n=17, 89\%), (b) probability of receiving the LSI (n=15, 79\%), and (c) patient information used by the decision-support model (n=15, 79\%). These three types of information were also ranked as the most important. Only a few participants thought that decision-support displays should present information about the model's development (n=3, 16\%) or performance (n=3, 16\%).

\subsection{Interviews: Designing Displays of Information Synthesis and Recommendations }

\subsubsection{Decision Support Design}
Based on the survey results and prior research on human-AI interaction strategies, we designed two decision-support displays: (a) a display with information synthesis and (b) a display with information and recommendations. To guide the selection of patient information for the displays (Figure \ref{fig:mockups}), we used variables included in Bayesian models predicting the risk of receiving blood transfusion and neurosurgical intervention during trauma resuscitation \cite{sullivan2023blood, sullivan2023nsi}. Inspired by Miller et al. \cite{miller2023explainable}, we explored different ways of representing patient data and different layouts (e.g., dividing the information pushing for an intervention and the information pushing against an intervention (Figure \ref{fig:mockups}.B)). 

\subsubsection{Interview Protocol and Data Analysis} To refine the design of the displays, we conducted remote semi-structured interviews with nine emergency medicine physicians (eight attending physicians and one fellow) from our primary research site (Appendix \ref{appendix_demopgrahics}). Their years of experience ranged from three to 35, with an average of 15 years. Because the earlier survey was anonymized, we do not know if interview participants had also completed the survey. All interview participants were offered compensation for their time.

\begin{figure*}
  \centering
  \includegraphics[width=1.0\textwidth]{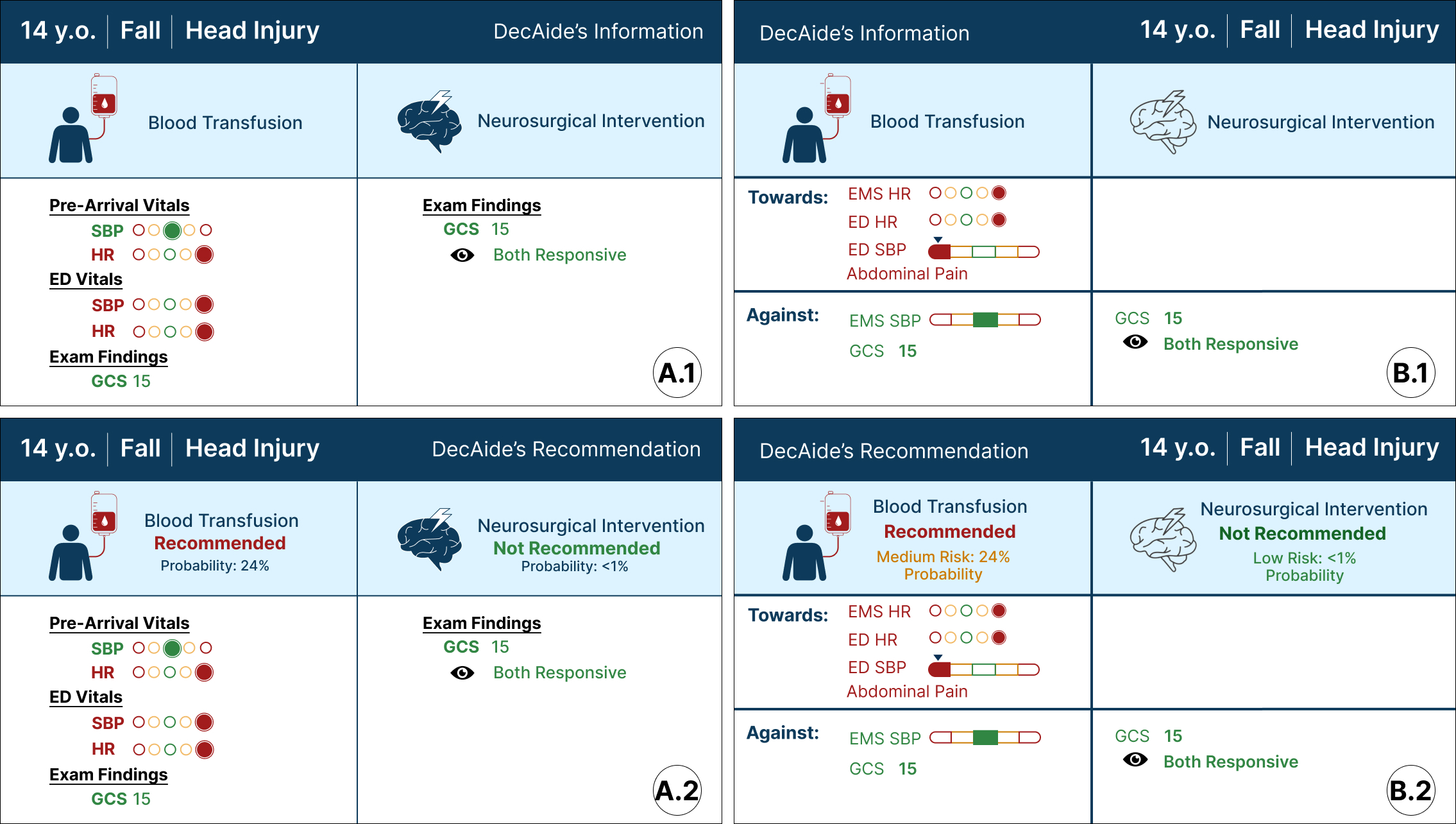}
  \caption{Examples of recommendation and information display mockups used to elicit participant feedback in the interviews. The top row has mockups of an information-synthesis display (A.1 and B.1) while the bottom row has mockups of an information-and-recommendation display (A.2 and B.2).}
  \Description{}
  \label{fig:mockups}
\end{figure*}

The interviews were 30 minutes long and split into the following sections: Introduction (3 minutes), Current Decision Making Processes (10 minutes),  Decision-Support Design (15 minutes) and Wrap-up (2 minutes). After introducing the interview study and obtaining consent for recording the interview, we asked participants to discuss their decision making process when evaluating if a patient has severe blood loss, the information they consider when determining if the patient requires a blood transfusion, and any opportunities for improving the decision making process. We then repeated these questions, focusing on traumatic brain injury and determining the need for neurosurgical intervention. Next, we conducted a 5-minute brainstorming activity, asking participants to sketch a display to support decision making about blood transfusion and neurosurgical interventions. We then showed mockups of our potential displays with different configurations of information and AI-enabled recommendations (Figure \ref{fig:mockups}). We used participant feedback to iteratively refine the mockups throughout the interviews.

All interviews were audio- and video-recorded with participant consent.  To maintain participant confidentiality and privacy, the interview recordings were stored on university-approved cloud servers and only accessible to members of the research team. We also removed identifiers from the auto-generated transcripts before beginning our analysis. An HCI researcher on the team used affinity diagramming \cite{moggridge2007designing} to inductively analyze the interview transcripts. We focused on identifying factors influencing the design of decision support for time-critical medical events. The findings and themes from the analysis were further discussed and refined through discussions with the entire research team, which includes HCI researchers and medical providers.

\subsubsection{Interview Findings: Receptiveness to AI support, but tension between providing clear advice and leaving room for clinical judgment.} Interview participants were open to using AI-enabled decision support during resuscitations. One participant described the use of AI during resuscitations as \textit{“inevitable”} [P5], while another noted that AI could \textit{“combine these many data points faster than I would as a human” }[P2]. The information synthesized in the display mockups matched the information participants said they often considered during resuscitations when assessing the need for blood transfusion and a neurosurgical intervention. Several participants [P1, P2, P3, P4, P5, P9] noted a preference for mockups that highlighted both abnormalities and temporal changes in dynamic data (e.g., A.1 and A.2 in Figure \ref{fig:mockups}). 

From participant feedback, we developed and followed three design guidelines for creating AI-enabled decision-support displays for time-critical medical events: (1) highlight abnormalities, (2) show temporal changes in dynamic data, and (3) provide a clean and concise design. Our final display designs included the number and severity of abnormalities to help providers quickly assess patient status and determine the need for interventions. Because patients with worsening vital signs may be more likely to require LSIs, we used color-coding to indicate changes in vital signs and help physicians determine the appropriate interventions. Finally, participants noted that displays should be clear, concise, and easy to interpret, given the dynamic and sometimes chaotic nature of resuscitations. We worked with clinical collaborators to ensure that the displays only included injured body regions relevant to determining the need for blood transfusion or neurosurgical intervention. For the displays with information synthesis and recommendations, participants noted the ability to use the information synthesis to \textit{"peek under the hood of the model"} [P6] and better understand the factors driving system recommendations.

The interview data also showed a tension between the desire for clear advice from decision support (e.g., recommendations) and the need to leave room for clinical judgment. The findings highlighted the trade-offs associated with providing information and recommendations or only information synthesis (without recommendations). Participants appreciated that the recommendations provided clear and unambiguous guidance on the next steps. For example, some participants found probability or risk levels hard to interpret on their own, preferring explicit recommendations: \textit{“If you put medium risk there, I'll be like, is it recommended, or is it not?”} [P8]. However, participants also worried about the potential issues that may arise when they disagreed with recommendations, stressing the ability to ignore recommendations they deemed inaccurate: 
\begin{quote}
    “[AI] is there to guide me and 
    to make me pause, and think maybe I should be considering things. But machines can be wrong too. They’re wrong less than humans, but they can still be wrong. So as long as my clinical judgment and the clinical judgment of my colleagues can be taken into account, I’m fine with it. It’s another piece of information…” [P7]
\end{quote}
Participants highlighted concerns about the legal ramifications of disagreeing with recommendations from the AI-enabled decision-support system, especially if risk predictions or recommendations were also documented in archival systems, such as electronic health records.

\section{Phase II: Evaluating AI-Enabled Decision Support}
In the second study phase, we evaluated the effectiveness of two human-AI interaction strategies -- (1) AI Information Synthesis and (2) AI Information and Recommendations -- through an online experiment with 35 providers from six hospital systems. This study phase was approved by our university's Institutional Review Board.

  \begin{figure*}
  \centering
  \includegraphics[width=1.0\textwidth]{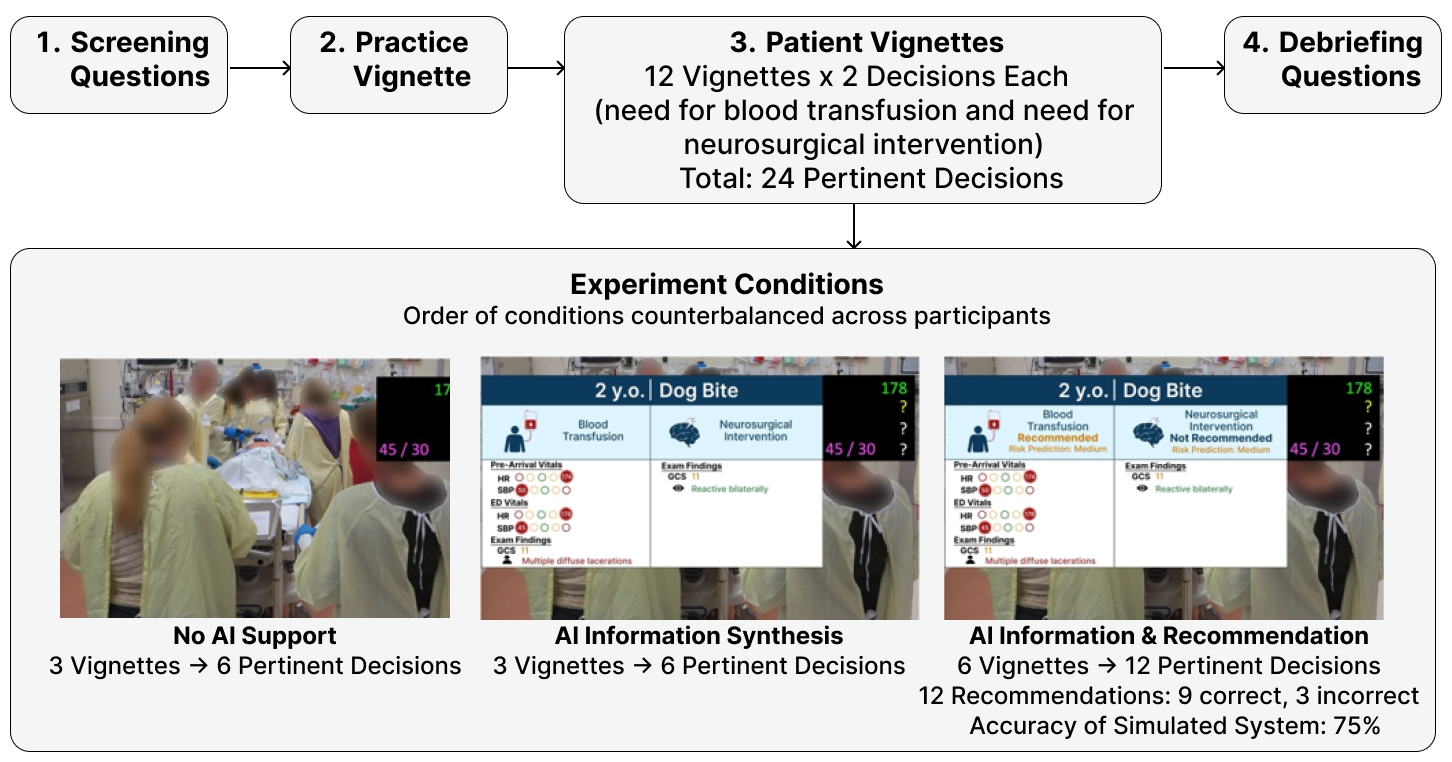}
  \caption{An overview of the online experiment flow. Participants experienced three conditions: (1) No AI Support, (2) Display with AI Information Synthesis, and (3) Display with AI Information and Recommendations. Participants were randomly split into groups, with the order of conditions counterbalanced across the groups. In the two conditions with a display, information was dynamically synthesized on the display as it was presented in the voiceover. In the condition with recommendations, recommendations and risk predictions for both blood transfusion and neurosurgical intervention appeared at the end.}
  \Description{}
  \label{fig:experiment_flow}
\end{figure*}

\subsection{Methods}
We conducted a within-subjects, online experiment with 35 providers. In the experiment, participants were presented with 12 simulated patient vignettes and asked to determine LSIs needed by each patient. The experiment was developed using Qualtrics (an online research platform) and consisted of four sections: (1) demographic and screening questions, (2) instructions and a practice vignette, (3) 12 patient vignettes, and (4) debriefing questions (Figure \ref{fig:experiment_flow}). Participants experienced three conditions during the experiment: (1) No AI Support, (2) Display with AI Information Synthesis, and (3) Display with AI Information and Recommendations. To mitigate learning and order effects, we randomly split participants into groups, with the order of conditions counterbalanced across the groups.

\begin{figure*}
  \centering
  \includegraphics[width=0.6\textwidth]{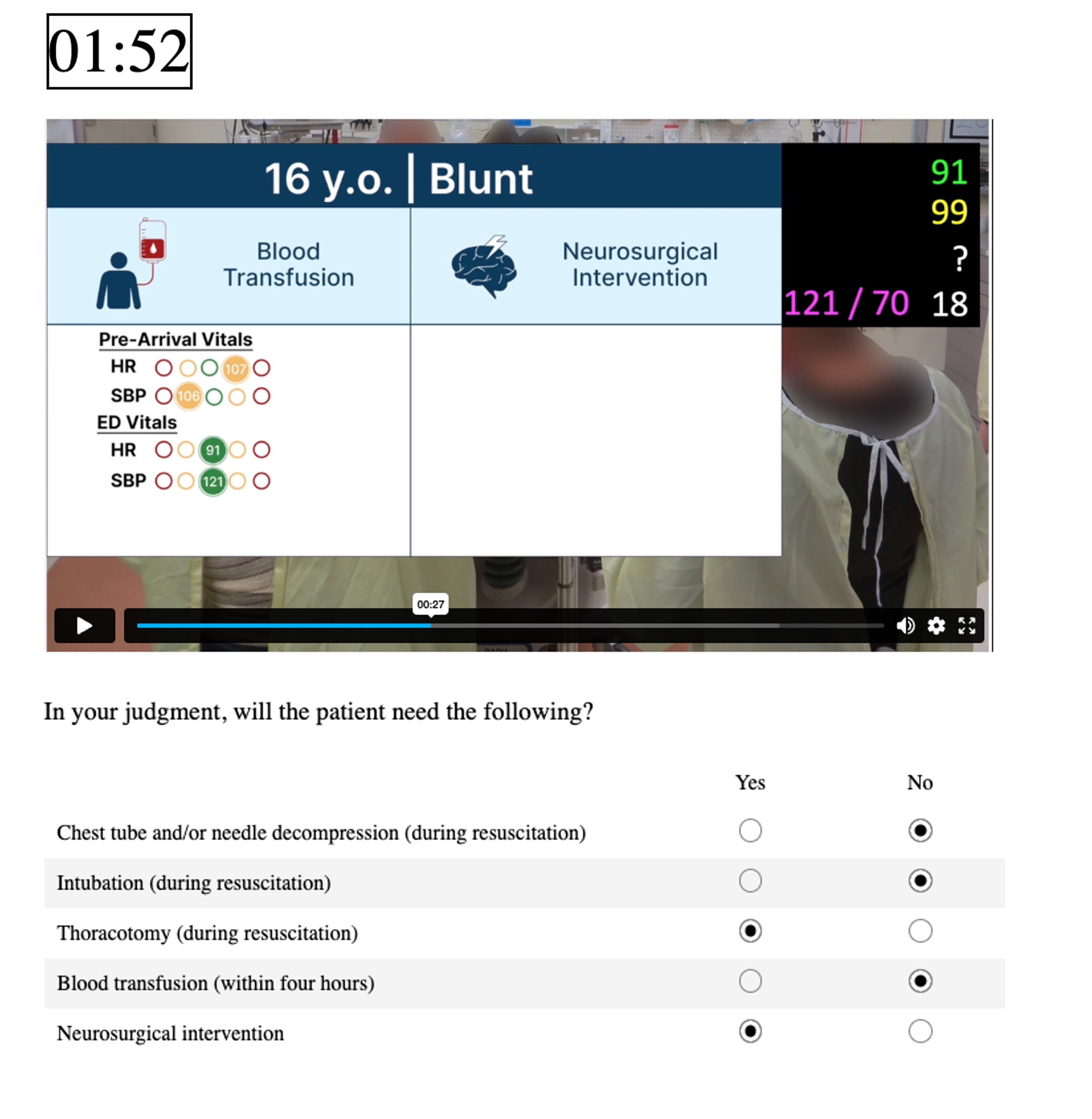}
  \caption{An example of a vignette included in the online experiment. The vignette contained a countdown clock and an embedded video with a voiceover of the case. In vignettes with AI Information Synthesis or AI Information and Recommendations, the video also included the AI support display. The questions about the patient's need for LSIs were located below the video.}
  \Description{}
  \label{fig:vignette}
\end{figure*}

\subsubsection{Vignette Selection and Ground Truth Data}\label{vignette_selection} We based the 12 patient vignettes on past resuscitations at our primary research site. When selecting the resuscitations, we focused on cases where decision making might have been challenging and could be most affected by decision support. We followed a data-driven approach to identify "borderline" cases using Bayesian models that had been developed to predict the risk of receiving blood transfusion or a neurosurgical intervention \cite{sullivan2023blood, sullivan2023nsi}. We first reviewed records of resuscitations from the past five years at our primary research site and obtained the patient information needed for the Bayesian models. We then used a perturbed tree discretization algorithm to classify the outputs of the Bayesian model into three categories: (1) low risk, (2) medium risk, and (3) high risk. We considered cases to be borderline if the probability of receiving a blood transfusion or a neurosurgical intervention was in the medium risk range or within one standard deviation outside the medium risk range. From this process, we identified 28 resuscitations that had a borderline risk for blood transfusion or neurosurgical intervention based on the Bayesian model. Our ground truth data were: (1) if the patient needed blood as determined by expert adjudication and (2) if the patient received a neurosurgical intervention. Because patients are often transfused unnecessarily or die from hemorrhage before receiving blood \cite{hess2005optimizing}, we adjudicated the need for blood transfusion in these potential cases. Two physicians, one from surgery and one from emergency medicine, independently reviewed the chart of each patient to determine the need for blood transfusion. Four potential cases were removed because of a disagreement between the two physicians. From the remaining 24 cases, we selected 12 for inclusion in the experiment: (1) six borderline cases for blood transfusion (three with the need for blood transfusion, three without) and (2) six borderline cases for neurosurgical intervention (three with the need for a neurosurgical intervention and three without). A physician on the research team used information from the medical record for the 12 cases to create the scripts for the voiceovers. These scripts were also reviewed for accuracy by an emergency medicine physician on the team.

\subsubsection{Simulated Patient Vignette Design} To design the simulated patient vignettes, we collaborated with clinicians on our team, ensuring the ecological validity of the experiment. We also conducted four pilots (two with medical students with experience in trauma resuscitation and two with advanced practice providers), iteratively resolving issues. The vignettes contained an embedded video with a voiceover of the simulated case, a vital signs monitor, and the AI support display depending on the condition (Figure \ref{fig:vignette}). The voiceover began with a simulated EMS report that includes the mechanism of injury, pre-hospital patient exam, and interventions. A medical provider then presented the key findings from the patient examination. We chose to present this information through a verbal report because trauma team leaders typically stand at the foot of the patient bed, receiving verbal reports from EMS and the bedside examiner.

In each vignette, the participant was asked to select if the patient needs certain LSIs, including (1) blood transfusion, (2) neurosurgical intervention, (3) a chest tube or needle decompression, (4) intubation, and (5) thoracotomy. Because our AI-enabled decision-support system focused on decisions about blood transfusion and neurosurgical intervention, we did not analyze diagnostic accuracy for the other three LSIs. We added these LSIs to increase the ecological validity of the experiment because trauma team leaders evaluate the need for multiple interventions during resuscitations. The length of the voiceovers ranged from 73 to 108 seconds, with an average of 91 seconds. To simulate the urgency and time pressure faced by resuscitation teams, we restricted the decision-making time. We limited the amount of time to select answers to the length of the voiceover plus 30 seconds. In our pilot, a 30-second time limit forced participants to make some decisions while still providing enough time to answer most questions. Each vignette had a countdown clock displaying the remaining time. The countdown clock started when the participant started playing the video.

\subsubsection{Simulated AI Decision-Support System} 
To control the accuracy of the recommendations and information synthesis, we simulated an AI decision support. Similar to how the system would function in actual medical events, the simulated system used automated approaches to capture vital sign measurements and relevant findings from the examination, showing these findings on the display. As information was shared in the voiceover, it was also dynamically added to the decision-support display in conditions with AI support. In the AI Information and Recommendations condition, risk predictions and recommendations appeared on the display at the end of the voiceover, after all information was presented. The risk predictions were based on the outputs of the Bayesian models, and the recommendations were based on ground truth data. 

Of the 12 vignettes, participants were shown three with No AI Support, three with AI Information, and six with AI Information and Recommendations (Figure \ref{fig:experiment_flow}). In the six vignettes with AI Information and Recommendations, each display had two recommendations, one for blood transfusion and the other for a neurosurgical intervention, resulting in 12 total recommendations. To investigate the effect of incorrect recommendations on decision making, some recommendations were intentionally incorrect. When determining the accuracy of the simulated system, we needed to balance having enough incorrect recommendations to evaluate over-reliance without causing participants to distrust the system. To achieve this balance, each participant received nine correct recommendations and three incorrect recommendations. As a result, our simulated system had an accuracy of 75\%, a level used in prior studies evaluating the effects of different AI support strategies on decision making \cite{buccinca2021trust, buccinca2024towards}. To make the simulated system mistakes appear realistic, incorrect recommendations were provided for interventions determined to be borderline by the Bayesian model predictions. At the beginning of the experiment, participants were informed that some vignettes may contain decision-support displays with AI information or AI information and recommendations. We told participants that the decision-support system could occasionally make mistakes but did not provide any specific information on the system accuracy to avoid biasing participants. Similarly, we did not provide any details about how the AI model used by the system was developed or validated.


\subsubsection{Debriefing Questions} After the 12 patient vignettes, we asked participants to rank the different types of AI support, describe how they used the information and recommendations (if at all), and state their perceived accuracy of the system. We also asked optional questions about potential design changes and any concerns about using the system in their practice (Appendix \ref{appendix}). 

\begin{table*}
  \caption{Overview of the participants in the evaluation phase of the study (n=35). The average years of experience (YOE) is provided for different demographic groups.}
  \label{tab:participants}
  \begin{tabular}{l|l|L}
    \toprule
    \textbf{Specialty}&\textbf{Number of Participants}&\textbf{Participant Demographics}\\
    \midrule
    Surgery&20&3 Attendings (Average YOE: 25)\newline 16 Residents (Average YOE: 4)\newline 1 Nurse Practitioner (YOE: 7)\\
    \midrule
    Emergency Medicine&10&6 Attendings (Average YOE: 17)\newline 4 Fellows (Average YOE: 4)\\
    \midrule
    Critical Care&5&3 Attendings (Average YOE: 19)\newline 2 Fellows (Average YOE: 6)\\
  \bottomrule
\end{tabular}
\end{table*}

\subsubsection{Participants} 
 We recruited physicians and advanced practice providers who had completed Advanced Trauma Life Support certification to ensure that participants were trained in making decisions about LSIs during trauma resuscitation. Following recruitment strategies from similar studies \cite{lee2023understanding, gu2023augmenting}, we distributed call-for-participation emails at our primary research site and through contacts of the research team. In addition to our primary site, we recruited participants from five other hospital systems in the United States. Participants were offered compensation for their time. To maintain participant anonymity, we collected the personal information needed for compensation (e.g., names and email addresses) in a separate survey. We did not link this information to the responses from the online experiment. 

 Thirty-five physicians and advanced practice providers completed the experiment and passed the attention checks (Table \ref{tab:participants}, full demographic information in Appendix \ref{appendix_demopgrahics}). Attention checks included playing the video for every vignette and providing logical answers to the debriefing questions. Because responses were anonymized, a breakdown of participants across the six hospital systems is not available. One critical care attending physician left a comment in the last debriefing question about having technical difficulties viewing the videos for two vignettes. As a result, we excluded data for this participant from all 12 vignettes but still considered their responses to the debriefing questions. No other participants reported technical issues that would require data exclusion.

\subsubsection{Data Points and Analyses} During the experiment, we collected the following data: (a) answers to the questions about the need for different LSIs and any changes made to the answers during the vignette, (b) times in the vignette when the participant selected if an intervention was or was not needed, (c) time when the participant finished the vignette, and (d) answers to the debriefing questions at the end of the experiment. We conducted the following analyses:
\begin{itemize}
  \item We used mixed-effects regression models to compare diagnostic accuracy (logistic regression), time to initial answer (linear regression), and time to final answer (linear regression) \cite{bates2015package}. Because the allocated decision-making time varied between the vignettes, time to initial and final answers was measured as a percentage of the allocated time for the vignette. Covariatesincluded (1) participant specialty (emergency medicine, surgery, or critical care), (2) if the participant was still in training (e.g., a resident or fellow), (3) the order in which the participant experienced different types of AI support, (4) the intervention being decided (blood transfusion or neurosurgical intervention), and (5) if the intervention had a borderline risk prediction score. We also initially included two random effects in the models: the participant and the vignette IDs. Including participant and vignette IDs as random effects would allow us to simultaneously account for variability within and across participants and vignettes \cite{brown2021introduction}. In the logistic regression model for diagnostic accuracy, participant ID explained zero variance in the data. We removed this random effect from the models related to diagnostic accuracy but retained it in other models. We conducted post-hoc pairwise comparisons using the R package \textit{emmeans} \cite{emmeans} with Holm-Bonferroni adjustments to account for multiple comparisons \cite{holm1979simple}. In all statistical tests, p<0.05 was considered a significant effect. We report odds ratio (OR) which indicates the strength of a statistical association and direction of effect (e.g., an OR greater than one indicates that a factor is associated with a higher likelihood of an event occurring) \cite{norton2018odds}. 
  \item We used the Friedman Test to compare participant rankings of the three types of AI support for effectiveness, efficiency, and overall preference. Post-hoc pairwise comparisons were performed using paired Wilcoxon signed-rank tests with Holm-Bonferroni corrections to account for multiple comparisons \cite{holm1979simple}.
  \item We used affinity diagramming to analyze the perception and uses of the three types of AI support and concerns about receiving AI support in clinical practice \cite{moggridge2007designing}. Findings from the affinity diagramming were discussed and refined with the larger research team, which included HCI researchers and medical providers.  
\end{itemize}

\begin{figure*}
  \centering
  \includegraphics[width=1\textwidth]{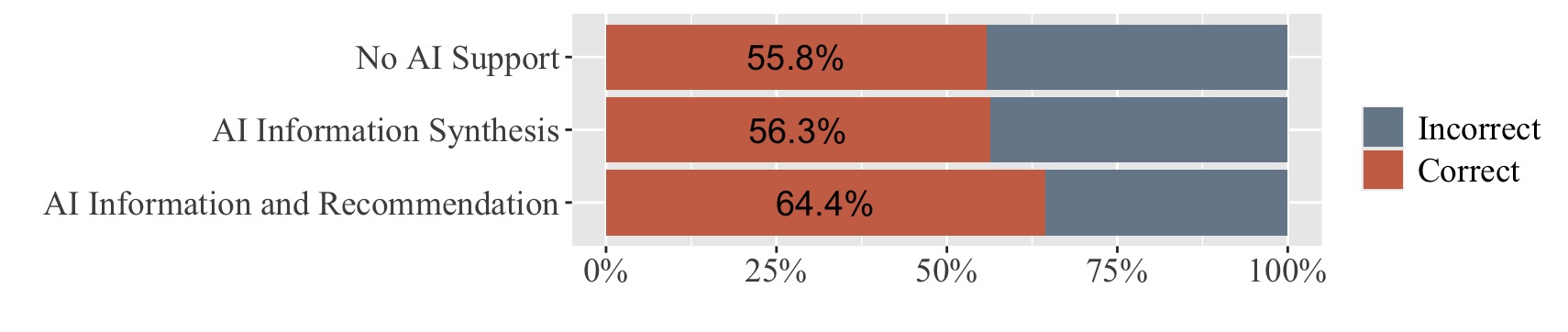}
  \caption{Diagnostic accuracy of decision-making instances across the three types of AI support. }
  \Description{}
  \label{fig:ai_support_summary}
\end{figure*}

\subsection{Findings}
In each vignette, participants selected whether the patient needed blood transfusion and/or neurosurgical intervention. We collected a total of 816 decision-making instances: 408 about blood transfusion and 408 about neurosurgical intervention. 

\begin{table*}
  \caption{Pairwise comparisons for diagnostic accuracy of all decision-making instances. We find support for H2: AI Information and Recommendations improves decision making compared to No AI Support.}
  \label{tab:accuracy}
  \begin{tabular}{l|c|c|c}
    \toprule
    Comparison&OR (95\% CI)&p-value&Hypothesis\\
    \midrule
    AI Information Synthesis-No AI Support&1.26 (0.79 - 1.98)& 0.33&H1\\
    \midrule
    AI Information \& Recommendation-No AI Support&1.80 (1.21 - 2.70)& 0.01*&H2 \checkmark\\
    \midrule
    AI Information \& Recommendation-AI Information&1.44 (0.96 - 2.14)&0.15&---\\
  \bottomrule
\end{tabular}
\end{table*}

\subsubsection{Effects of Decision Support on Diagnostic Accuracy}
The type of AI support (No AI Support, AI Information Synthesis, or AI Information and Recommendations) had a significant association with diagnostic accuracy in the mixed-effect logistic regression model (\begin{math}\mathcal{X}^2(2, N=816) = 9.06, p=0.01\end{math}; Figure \ref{fig:ai_support_summary}). In the post-hoc comparisons, decisions were more likely to be correct when AI Information and Recommendations were provided than when No AI Support was provided (OR=1.80, 95\% Confidence Interval (CI) for OR [1.21, 2.70], \textit{z} = 2.88, p=0.01; Table \ref{tab:accuracy}). We did not find a difference in diagnostic accuracy between AI Information Synthesis and No AI Support (\textit{z} = 0.98, p=0.33) or AI Information and Recommendations and AI Information Synthesis (\textit{z} = 1.78, p=0.15). Three covariates were also associated with diagnostic accuracy in the mixed-effects logistic regression model. Decisions without a "borderline" risk prediction score (OR=5.76, 95\% CI [3.67, 9.02], \begin{math}\mathcal{X}^2(1, N=816) = 69.52\end{math}, p<0.001) and decisions about neurosurgical intervention (OR=1.74, 95\% CI [1.18, 2.57], \begin{math}\mathcal{X}^2(1, N=816) = 8.12\end{math}, p<0.01) were more likely to be correct. Specialty was associated with diagnostic accuracy (\begin{math}\mathcal{X}^2(2, N=816) = 7.44\end{math}, p=0.02), with pairwise comparisons showing critical care physicians more likely to correctly determine if an intervention was needed than emergency medicine physicians (OR=2.22, 95\% CI [1.16, 4.27], \textit{z}= 2.42, p=0.046). In a follow-up analysis, we examined the effect of correct versus incorrect recommendations, focusing only on the instances with AI recommendations. We did not find that the recommendation correctness had an effect on diagnostic accuracy (\begin{math}\mathcal{X}^2(1, N=408) = 0.52\end{math}, p=0.47). Because the risk prediction score was associated with diagnostic accuracy and we expected to see the greatest effect of AI support on instances with "borderline" risk predication scores, we conducted another follow-up analysis of these instances only. The type of AI support again had a significant association with diagnostic accuracy (\begin{math}\mathcal{X}^2(2, N=510) = 8.28, p=0.02\end{math}). Decisions were more likely to be correct when AI Information and Recommendations were provided as compared to No AI Support (OR=2.23, 95\% CI [1.29, 3.87], \textit{z} = 2.85, p=0.01). No other pairwise comparisons for AI support were significant.

\subsubsection{Timing of Information, Recommendations, and Decisions}
\begin{figure*}
  \centering
  \includegraphics[width=1\textwidth]{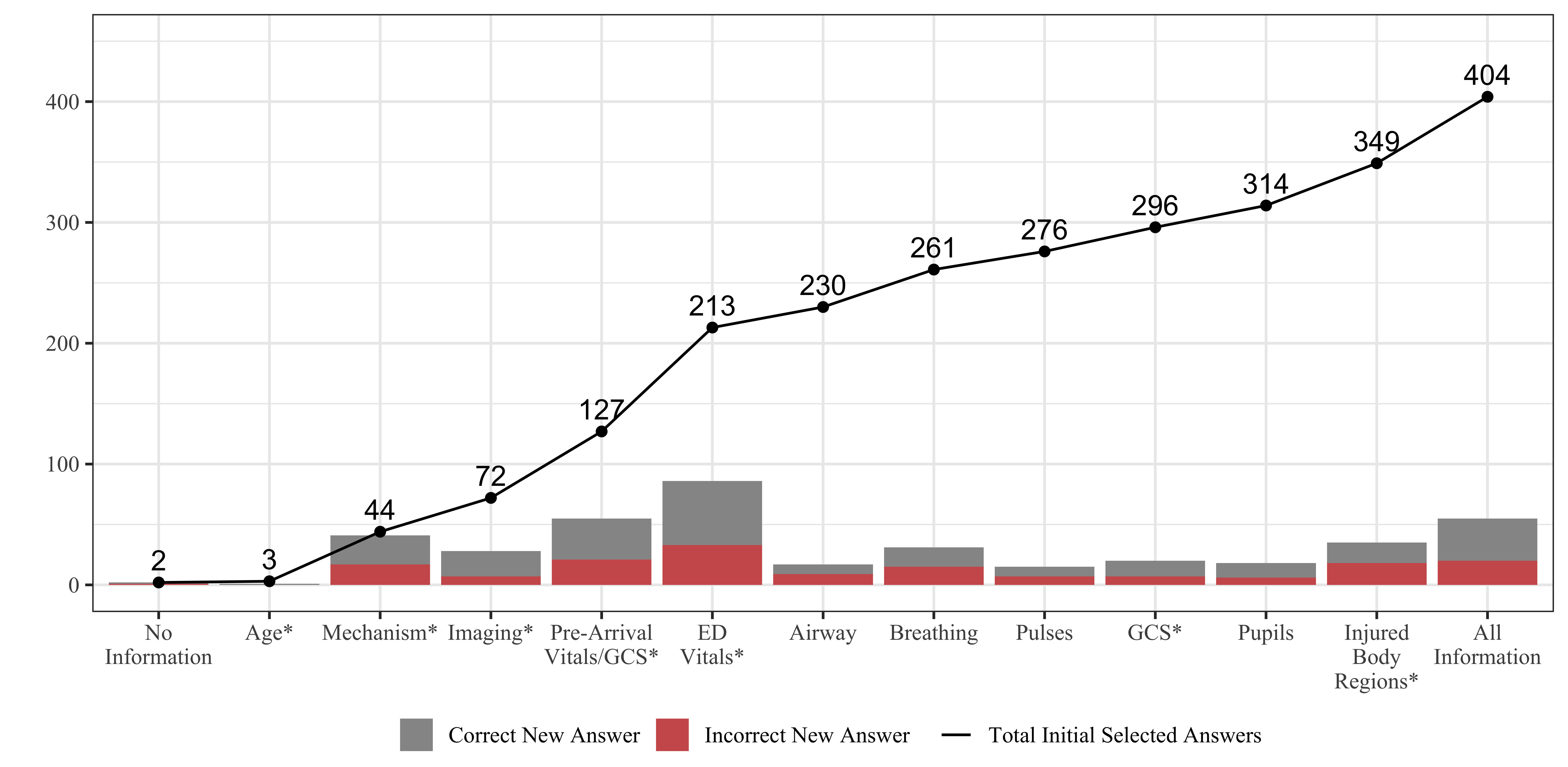}
  \caption{Timing of initial answers to the need for blood transfusion question. Information was presented in the same order for each vignette. The bar graph shows the number of new answers selected after a piece of information was presented (e.g., 41 new answers were selected after the mechanism of injury was presented). The bar graph also highlights the proportion of correct (gray) and incorrect (red) new answers. The line graph indicates the number of total initial answers (e.g., 44 total answers had been made from the start of the video to after presentation of mechanism of injury). An asterisk (*) indicates that the piece of information was also included on the decision-support display.}
  \Description{}
  \label{fig:blood_timings}
\end{figure*}

\begin{figure*}
  \centering
  \includegraphics[width=1\textwidth]{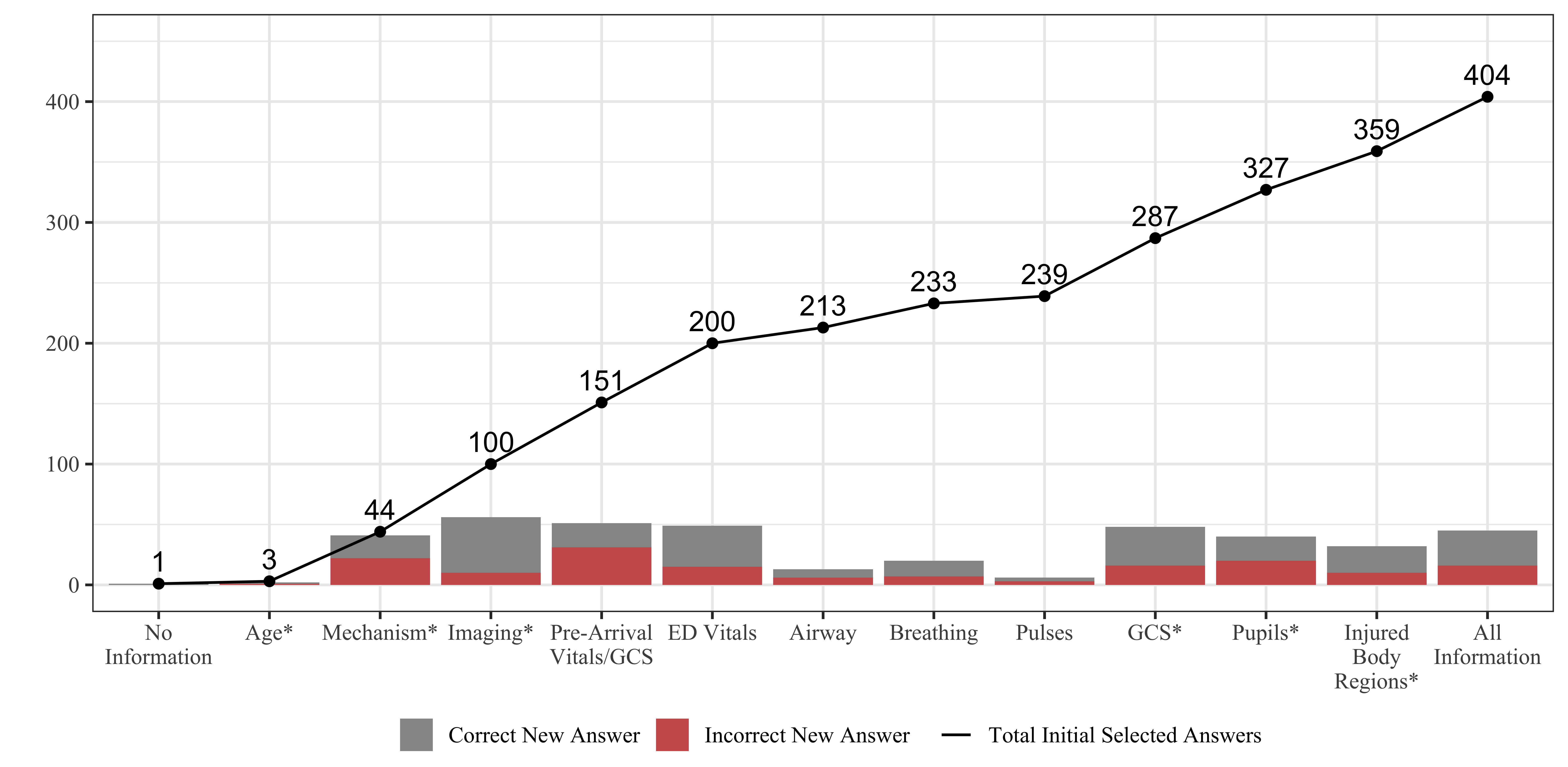}
  \caption{Timing of initial answers to the need for neurosurgical intervention question. This visualization follows the same structure as Figure \ref{fig:blood_timings}.}
  \Description{}
  \label{fig:tbi_timings}
\end{figure*}

We did not find an association between the type of AI support and the time taken to select an initial (\begin{math}F_{2,768.2} = 0.14, p=0.87\end{math}) or final (\begin{math}F_{2,768.3} = 1.07, p=0.34\end{math}) decision. Information about the case was reported in the same order in all voiceovers, following the typical structure of EMS reports and task order in the ATLS protocol \cite{subcommittee}. Participants often made their initial decisions about the need for blood transfusion or a neurosurgical intervention before all information was presented (Figures \ref{fig:blood_timings} and \ref{fig:tbi_timings}). In about half of the total decision-making instances for both LSIs, an initial answer was selected after the pre-arrival information and emergency department vital signs were presented, but before information about the primary and secondary survey findings. Participants changed their initial answers in 89 instances (11\%), sometimes changing an answer to the same question multiple times. Answers were more frequently changed for decisions about neurosurgical intervention (n=57) than for decisions about blood transfusion (n=32). Of the 816 decision-making instances, we only observed eight instances where the participants ran out of time before selecting an answer.

\begin{figure*}
  \centering
  \includegraphics[width=1\textwidth,trim=4 4 4 4,clip]{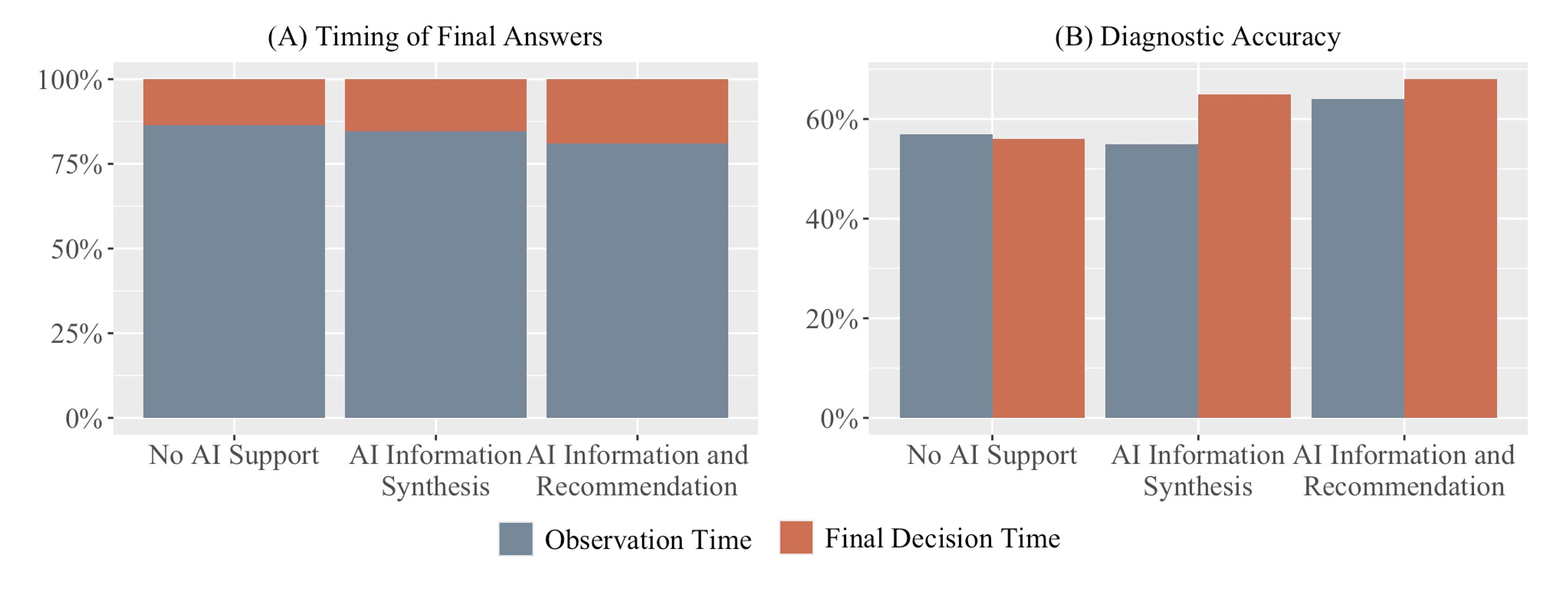}
  \caption{(A) The proportion of answers selected during the observation and final decision periods for the three AI support conditions. (B) The diagnostic accuracy level of the answers selected during the observation and final decision periods for the three AI support conditions.}
  \Description{}
  \label{fig:observation_decision_time}
\end{figure*}

We also examined the time in the vignette when the final answer was selected for all decision-making instances. The initial answer time and final answer time differed in instances where the participant changed their first answer. The two times were the same in instances where the participant did not change their answer. In this analysis, we split the overall time spent on a vignette into two periods: the observation and final decision period \cite{cao2023time}. The observation period was the time spent listening to the voiceover. During this period, new information about the simulated patient was provided to the participant. The decision period was the period between the end of the voiceover and the time when the participant moved to the next vignette. During this period, new information was shown in only the AI Information and Recommendations condition and included the risk prediction and recommendation. The other two conditions had no new information presented during the decision period. Most final answers were selected during the observation period, while information was still being presented (Figure \ref{fig:observation_decision_time}). In the vignettes with No AI Support, final answers selected during the observation and final decision periods had similar diagnostic accuracy (56\% and 55\% respectively). In the vignettes with AI Information Synthesis, final answers selected during the final decision period had higher diagnostic accuracy (65\%) than answers selected during the observation period (55\%). When AI Information and Recommendations were provided, final answers selected during the observation period were accurate 64\% of the time and final answers selected during the decision period were accurate 68\% of the time.

\begin{figure*}
  \centering
  \includegraphics[width=1.0\textwidth,trim=4 4 4 4,clip]{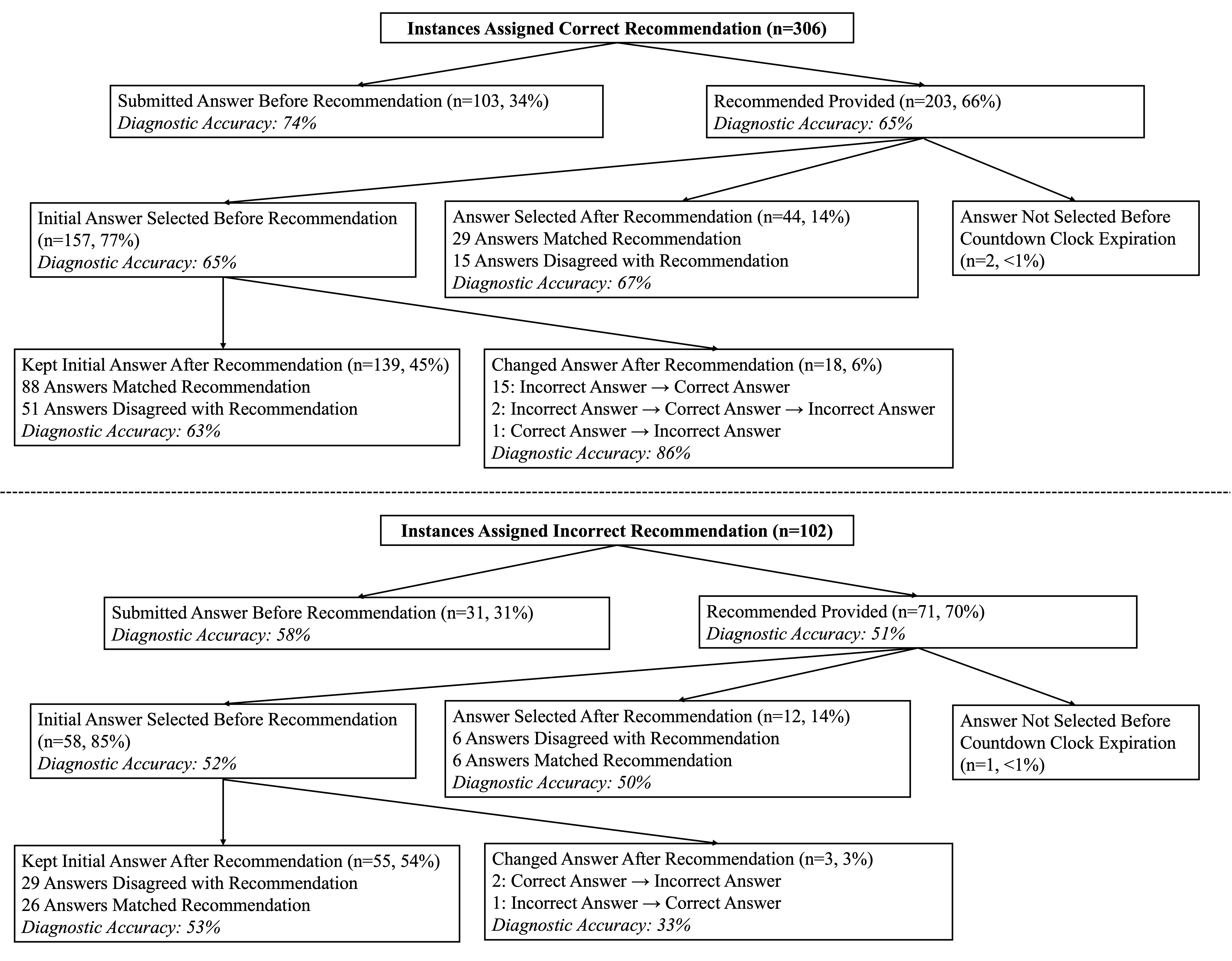}
  \caption{A breakdown of answers, decision-making behaviors, and diagnostic accuracy in the vignettes with AI Information and Recommendations. The top chart shows instances with correct recommendations and the bottom chart shows instances with incorrect recommendations. Arrows within boxes indicate answer change.}
  \Description{}
  \label{fig:answer_breakdown}
\end{figure*}

Recommendations appeared on the decision-support display at the end of the voiceover in vignettes with AI Information and Recommendations. In 103 of 306 decision-making instances assigned to displays with a correct recommendation, participants had selected an answer and moved to the next vignette before the recommendations were shown (Figure \ref{fig:answer_breakdown}). In the remaining 203 instances with a correct recommendation, we observed 132 instances where the final answer matched with the correct recommendation. Participants selected the correct answer before the recommendation appeared (n=88), switched from an incorrect to correct answer after the recommendation appeared (n=15), or selected an initial and correct answer after the recommendation appeared (n=29). We also observed 69 instances where the participant submitted an incorrect answer after a correct recommendation appeared on the display. Participants selected the incorrect answer before the recommendation appeared (n=51), switched to an incorrect answer after the recommendation appeared (n=3), or selected an initial (and incorrect) answer after the recommendation appeared (n=15). In two instances with a correct recommendation, the participant did not make a decision before the countdown clock expired.

In 31 of 102 decision-making instances assigned to displays with an incorrect recommendation, participants had selected an answer and moved to next vignette before recommendations were shown (Figure \ref{fig:answer_breakdown}). In the remaining 71 instances, the final answer disagreed with the incorrect recommendation in 36 instances. Participants selected the correct answer before the recommendation was shown and did not change this answer (n=29), switched from an incorrect to correct answer after the incorrect recommendation appeared (n=1), or selected an initial and correct answer after the incorrect recommendation appeared (n=6). We also observed 34 instances where the participant submitted an incorrect answer after the incorrect recommendation appeared. Participants had selected the incorrect answer before the incorrect recommendation appeared (n=26), switched to an incorrect answer after the recommendation appeared (n=2), or selected an initial and incorrect answer after the recommendation appeared (n=6). In one instance with an incorrect recommendation the participant did not make a decision before the countdown clock expired.

\subsubsection{Rankings of Effectiveness, Efficiency, and Overall Preference}
\begin{figure*}
  \centering
  \includegraphics[width=1.0\textwidth]{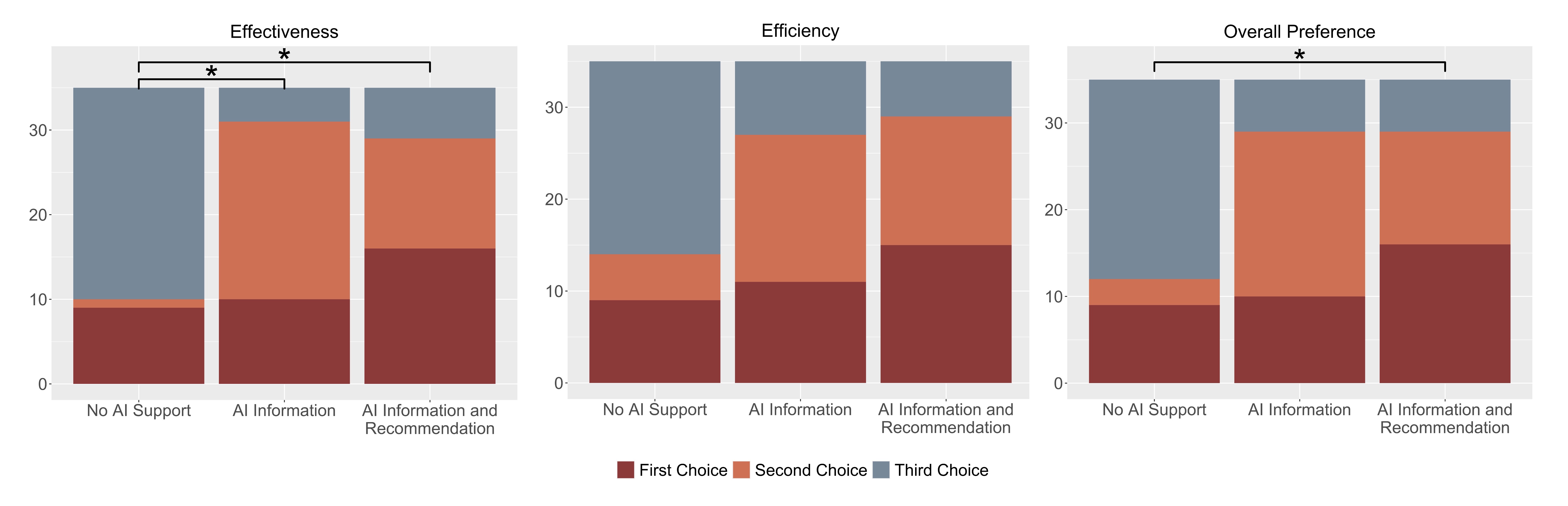}
  \caption{Participant rankings of the three AI support types for effectiveness, efficiency, and overall preference. An asterisk (*) indicates a significant difference between pairs.}
  \Description{}
  \label{fig:rankings}
\end{figure*}

We found differences in participant rankings of the three AI support types for effectiveness (\begin{math}\mathcal{X}^2(2, N=35) = 11.20, p<0.01\end{math}), efficiency (\begin{math}\mathcal{X}^2(2, N=35) = 6.70, p=0.04\end{math}), and overall preference (\begin{math}\mathcal{X}^2(2, N=35) = 8.91, p=0.01\end{math}) (Figure \ref{fig:rankings}). In the post-hoc pairwise comparisons for effectiveness, participants ranked AI Information Synthesis (\textit{z} = 2.39, p=0.05) and AI Information and Recommendations (\textit{z} = 2.57, p=0.03) as more effective at supporting decision making than No AI Support. More participants preferred AI Information and Recommendations over No AI Support (\textit{z} = 2.40, p=0.049). No pairwise comparisons between the three types of AI support were significant for the efficiency rankings. 

\subsubsection{Perceptions and Uses of Different Types of AI Support}
From our analysis of participant answers to the debriefing questions, we found that participants had polarizing perceptions and uses of the AI recommendations. Eighteen participants indicated that they had considered recommendations from the system during decision making, noting that recommendations often arrived after they had made an initial decision. When the recommendation did not align with their initial decision, several participants discussed being prompted to reflect on their decision:
\begin{quote}
    "Mostly [the recommendation] was confirming what I planned already. But at times it was divergent from my thinking and made me step back and consider why the AI was recommending something. This is, in itself, useful even if AI is 'wrong' because it forces me to stop and think." [P3 - Emergency Medicine Attending] 
\end{quote}
In contrast, one participant found it \textit{"discomfiting"} [P18 - Surgical Attending] to receive a recommendation that differed from their decision, noting that it prompted \textit{"second guessing"}. An emergency medicine attending described changing some of their decisions after receiving recommendations but not always being sure if the changes were correct.

Twelve participants ignored the AI recommendations. Several of these participants thought the recommendations lacked nuance, noting that multiple factors can influence decisions during medical emergencies like trauma resuscitation. Others did not trust the system because the recommendations could be incorrect and no information was provided about the process used to train the system:
\begin{quote}
    "[...] Many of these decisions are based on values and not data. How is the AI engine trained? Whose recommendations is it using? I know how my colleagues are trained and it gives me confidence to trust them. I don't have that same confidence in the AI system." [P18 - Surgical Attending]
\end{quote}

When asked if they had any concerns about using a system with AI recommendations in clinical practice, participants frequently mentioned bias and over-reliance on the recommendations as potential issues. Participants also had concerns about their ability to ignore recommendations from the AI system, worried that proceeding with a different plan could cause distrust among team members or lead to legal consequences. One participant highlighted that ignoring the recommendations of the AI system could potentially be more challenging for junior physicians:
\begin{quote}
    "I think we quickly start to anchor on the AI unconsciously. As a junior trainee, I feel like I was more 'learning from the AI' than the other way around. In real-time, I may be more afraid to go against what the AI says because I assume the AI was trained on the best standard but I have no idea how accurate it is" [P28 - Surgical Resident]
\end{quote}
However, another resident thought that ignoring recommendations would be hard but possible, noting that physicians have experience disagreeing with the results of diagnostic tests and \textit{"taking them in stride with [their] clinical judgment"} [P35 - Surgical Resident]. 

Participants described using the AI information synthesis to review findings from patient examinations, identify abnormalities, and check their logic. Participants noted that verbally-reported information could be challenging to track during the resuscitation, appreciating the ability to review reported information with AI information synthesis. Participants also appreciated that the information synthesis indicated the normality of vital signs. One participant described using the information synthesis to \textit{"check [their] logic when listening to the case" [P22 - Critical Care Fellow]"}. Overall, participants had fewer concerns about receiving AI information synthesis during real practice. Compared to the recommendations, participants thought that AI information synthesis should be \textit{"pretty objective"} [P28 - Surgical Resident]. However, a few participants did raise concerns that the information could be incorrect, incomplete, or misleading. One participant highlighted that the information synthesis could put \textit{"emphasis on the wrong piece of information or de-emphasize information that is not given"} [P25 - Surgical Resident].

\section{Discussion}
In this study, we examined the effectiveness of information-driven support and the role of recommendations through the design and evaluation of an AI-enabled decision-support system within the context of pediatric trauma resuscitation. After conducting design research with trauma team leaders, we developed two AI support displays: (1) a display with AI information synthesis and (2) a display with AI information and recommendations. We then evaluated the effects of these two human-AI interaction strategies on decision making in an experiment with 35 domain experts. Supporting our second hypothesis, we found that AI information and recommendations significantly improved diagnostic accuracy compared to no AI support. In contrast, we did not find a significant difference in diagnostic accuracy between only AI information synthesis and no AI support. No  difference in time taken to make a decision was found between the three conditions, suggesting that the improved accuracy with AI information and recommendations did not come with a trade-off of taking longer to make decisions. 

\subsection{Human-AI Interaction During Medical Emergencies}
Our findings from the experiment indicate that the AI support displays may have functioned as a type of process-oriented support, where information is continuously provided to users to help in reasoning about a problem \cite{zhang2024beyond}. Participants often made initial decisions as information about the patient was presented and synthesized on the display. Participants described using the synthesis to review pertinent exam findings, identify abnormalities, and check their logic. Additionally, participants had often made initial decisions before receiving recommendations. As a result, the recommendations frequently served as a cognitive forcing function \cite{buccinca2021trust}, with participants describing using the recommendations to check their initial decisions. 

Process-oriented support may be especially well-suited for medical emergencies. AI-enabled decision-support systems have often been designed for clinical settings that rely on image interpretation, such as radiology and pathology \cite{corti2024moving, gu2023augmenting, fogliato2022goes, thieme2024challenges, yildirim2024multimodal,cai2019human}. AI systems used in these settings can pre-compute decision-support recommendations, providing recommendations to providers when they begin viewing the images. However, pre-computing AI support is more challenging in emergency medical scenarios. Patient information is continuously obtained throughout the event, and the AI system will likely receive most information at the same time as the medical team (or potentially later). By following a process-oriented support strategy, AI systems can facilitate decision making by synthesizing information as it is obtained by the medical team, triggering recommendations after a sufficient amount of information has been captured. 

\subsection{Role of Recommendations in Information-Driven Strategies}
Our results suggest that including recommendations in information-driven support can be beneficial. Unlike recommendation-driven strategies, information-driven strategies do not center on providing recommendations \cite{gajos2022people, miller2023explainable, zhang2024beyond}. One open research area is understanding whether recommendations still have a role in this new human-AI interaction paradigm. Providing information synthesis and recommendations improved the decision making of our study participants compared to no AI support. Providing only information synthesis did not significantly affect diagnostic accuracy compared to no AI support. 

The displays of only AI information synthesis may not have challenged participants to consider other decisions in the same way that receiving a conflicting recommendation challenges decision making. In time-critical decision making, providers may look for and prioritize evidence that supports their initial decisions instead of seeking evidence that challenges these decisions (e.g., anchoring and confirmation biases) \cite{croskerry2013cognitive, doherty2020believing}. Recent work has proposed extending human-AI collaboration frameworks to have AI play the role of ``provocateur'', challenging user perceptions and ideas to improve decision making \cite{sarkar2024ai}. In our experiment, conflicting recommendations often served as provocateurs, challenging initial decisions that were made with limited information and may have been influenced by cognitive biases. However, our study only evaluated one type of information synthesis out of many. Other types of information synthesis may prompt critical thinking and reflection more effectively, making recommendations unnecessary. 
 
Although providing AI information and recommendations improved diagnostic accuracy, we still observed instances where participants received correct recommendations but made incorrect decisions. Several factors could have contributed to these results. First, prior work has found that receiving recommendations after an initial decision can reduce acceptance of recommendations~\cite{buccinca2021trust, fogliato2022goes, swaroop2024accuracy}. Second, because our participants were experts trained in making decisions about LSIs, they may have been less likely to rely on the output of AI-enabled decision-support systems than non-experts. Third, participants frequently referenced concerns about over-relying on the system, noting that it could provide incorrect recommendations. As a result, participants might have been hesitant to accept recommendations from the system. This finding aligns with prior work, which found that providers rated the quality of advice from other providers more poorly when it was labeled as AI generated \cite{gaube2021ai}.

\subsection{Barriers to Implementing AI Recommendations for Time-Critical Decision Making}
Although providing AI information and recommendations improved diagnostic accuracy in our experiment, our findings also highlighted two socio-technical barriers to implementing AI recommendations in time-critical medical events.

\subsubsection{Determining When to Provide Decision Support}
One challenge in implementing AI recommendations for time-critical decision making is determining exactly when to provide recommendations. Process-oriented support proposes that recommendations may be provided at later points in events to confirm or challenge the initial decisions of users \cite{zhang2024beyond}. As in this study and prior work \cite{croskerry2002achieving}, providers make decisions at different points in medical emergencies. Similar to the accuracy-time trade-off in decision making \cite{swaroop2024accuracy}, AI systems used during medical emergencies will also face an accuracy-time trade-off. Recommendations made by the system earlier in the medical event may be more likely to affect decision making and prevent teams from pursuing incorrect goals (e.g., preparing blood transfusion for a patient that does not need one). However, these recommendations may have lower accuracy because they are generated with limited information. Recommendations made later in the event with more information may be more accurate but less likely to affect decision making and the goals pursued by the team. Finally, if delays in provider decision making are leading to delays in intervention performance and patient deterioration, recommendations may need to be provided before initial decisions are made by providers. 

\subsubsection{Effective but Polarizing: Supporting Multiple Human-Centric Objectives}
In addition to improving decision making, AI systems may need to support multiple human-centric objectives, such as learning, collaboration, and agency \cite{buccinca2024towards}. Although providing recommendations supported decision making in our experiment, our findings suggest considering potential negative effects of AI recommendations on collaboration and agency during time-critical medical events. Our experiment evaluated how individual providers used AI information and recommendations and the effects of the recommendations on their decision making. However, decisions are often made collaboratively in high-stakes domains, such as healthcare \cite{kaltenhauser2020you} and aviation \cite{zhang2024beyond}. During medical emergencies, fast-response teams frequently include physicians from different specialties \cite{ford2016leadership}. Prior research has suggested that decision support could function as a coordination mechanism, aligning physicians from different specialties and experience levels on a treatment plan \cite{mastrianni2022alerts}. Participants in our design research phase also highlighted that decision-support displays should be viewed by the entire team to support a shared mental model. Although physicians envisioned decision support as a potential coordination mechanism, conflicting perceptions and uses of AI recommendations among domain experts may impede this goal. For example, pilots had diverging views of the usefulness of recommendations in a study comparing recommendation-driven and information-driven support \cite{zhang2024beyond}. Similarly, we found that providers in trauma resuscitation had polarizing perceptions and uses of AI recommendations, with some participants raising concerns that AI recommendations could lead to conflict and distrust among team members.

Another challenge in designing AI support for clinical settings is prompting critical thinking or ``provocation'' without challenging provider expertise \cite{zhang2024rethinking}. Some participants appreciated being prompted to stop and think about recommendations that conflicted with their initial decisions. Others expressed concerns about their agency to ignore recommendations. Because providers are legally liable for their decisions, agency, autonomy, and authority are important factors when considering the ethical implications of designing AI support for medical events \cite{morley2020ethics}. Agency, autonomy, and authority could be further reduced if AI recommendations are included in archival systems that can be used outside the medical event (e.g., as evidence in a medical malpractice lawsuit). Concerns about provider liability have been a long-recognized challenge in recommendation-driven AI assistance. If there is any institutional friction to disagreeing with AI recommendations, such as increased chance of liability for a medical error resulting from overriding AI recommendation, the presence of the algorithmic recommendation can be seen as taking away some authority from the human decision-maker without assuming any of the responsibility for the decision~\cite{woods1986paradigms,green2022flaws,kawakami2022improving}.

\subsection{Implications for Developing AI-Enabled Decision Support for Time-Critical Events}
From our findings, we propose three implications for developing AI-enabled decision support for time-critical events: 
\begin{enumerate}
    \item \textit{Support human decision makers in critically evaluating AI-generated information synthesis.} Our findings highlight that providers may be more critical of recommendations than information synthesis, suggesting a need to promote critical evaluation of information synthesis when developing AI-enabled decision support. Participants often raised concerns about over-relying on AI recommendations and making biased or incorrect decisions. However, few participants had concerns about receiving information synthesis, which was considered objective. Despite this perception, information synthesis can also bias decision making by emphasizing misleading, incorrect, or unimportant information \cite{buccinca2024towards, morrison2024impact}. Because users may evaluate information synthesis less critically than recommendations, information-driven strategies may need to prompt critical evaluation of the presented syntheses. One potential design strategy for prompting critical evaluation could involve highlighting any missing information in addition to synthesizing obtained information. For example, our displays showed the normality of patient information collected by the team but omitted potentially critical information that was not yet collected. Displaying the missing information might have prompted providers to assess the trade-offs of waiting for more information before making a decision.
    \item \textit{Consider information-driven strategies, beyond only recommendations, to support providers in managing accuracy-time trade-offs.} 
    Prior work assessing human-AI collaboration under time pressure has focused on recommendation-centric strategies \cite{swaroop2024accuracy, rastogi2022deciding, cao2023time}. Providing correct recommendations could assist providers in making correct decisions more quickly \cite{swaroop2024accuracy}. However, other types of information-driven AI support may also assist human decision makers in managing accuracy-time trade-offs, especially during dynamic events. During time-critical events, human decision-makers often have to weigh the trade-offs of waiting for additional information to make a more informed decision. For example, one provider described how they may assess the trade-offs of waiting for a CT scan to perform neurosurgical intervention. AI systems could provide information on how likely a piece of information is to change a recommendation, how long that piece of information will take to obtain, and how patient status might change during that time. AI systems could also help decision makers prioritize the order in which they obtain information. These forms of information-driven support could aid human decision makers in assessing the accuracy-time trade-offs of their decisions.
    \item \textit{Develop clear policies around the responsibility and liability of human decision makers when adopting AI-enabled tools.} Similar to findings from prior work, our participants raised concerns about the effects of AI-enabled tools on their authority and liability \cite{woods1986paradigms,green2022flaws,kawakami2022improving}. Institutions adopting AI-enabled tools may need to develop clear policies that define the authority and legal protections of human decision makers. Ideally, these policies would be developed with input from decision makers at the institution. In team-based settings, policies may need to distinguish how the AI-enabled tools impact the authority, responsibility, and liability of different team members.  In addition to providing information on system development, training, and accuracy in system onboarding sessions \cite{cai2019hello}, these sessions could also highlight the policies around responsibility and liability.
\end{enumerate}

Although we developed these implications from our research on decision support for trauma resuscitation, they may also apply to other time-critical events, such as firefighting and aviation \cite{zhang2024beyond, zhang2024exploring}. In these contexts, teams frequently make decisions with limited information under time pressure. Information-driven strategies could support decision makers in assessing the trade-offs of waiting for additional information and understanding what information to prioritize. In addition, these settings are often team-based and could also benefit from clear policies describing the impacts of AI-enabled tools on the responsibilities and liabilities of different team roles.

\subsection{Study Limitations and Opportunities for Future Work}
Our study has four main limitations. First, the design research phase included only participants from our primary research site and most were emergency medicine physicians. Although our design research participants were limited, our evaluation phase involved physicians from three specialties and six hospital systems. Second, only five critical care physicians participated in the evaluation phase of the study, compared to 20 surgical and 10 emergency medicine specialists. Future research with more critical care physicians is needed to further investigate the differences in decision making between different specialties involved in treating patients during medical emergencies. In addition, our sample size was relatively small due to the difficulty and cost in recruiting physicians and surgeons. However, this sample size is in line with similar studies that evaluated human-AI interaction with domain experts \cite{panigutti2022understanding,gu2023augmenting,lee2023understanding,zhang2024beyond,sivaraman2023ignore}. Although only 35 participants completed the experiment, we collected a total of 816 decision-making instances. Third, to evaluate the effectiveness of different human-AI interaction strategies, we used an online experiment instead of in-person simulations. The fidelity of the online experiment was ensured through pilot testing and continuous input from clinicians on our research team. Although we had challenges in recruiting clinical experts for the online experiment, recruiting for in-person studies is even more challenging because providers need to be in the hospital to attend the simulation session. Online experiments are valuable because they can provide initial insights into the effectiveness of different human-AI interaction strategies before conducting in-person simulations, which are more resource- and time-intensive. Because different hospitals may have different approaches to treating patients with life-threatening injuries, conducting the online experiment with participants from five other hospital systems outside of our primary research site allowed us to collect a wider set of perspectives. Our fourth study limitation is that all hospitals were level 1 trauma centers, which provide the highest levels of care. Further research is needed to understand the effects of human-AI interaction strategies during time-critical medical emergencies at other types of hospitals, such as those in rural areas or with fewer resources. Because our study examines how individual providers interacted with AI support, future work should also study how teams of providers interact with AI support, the effects on shared decision making, and the ethical implications of different types of AI support.

\section{Conclusion}
We designed and evaluated an AI-enabled decision-support system to support providers in making decisions about the need for life-saving interventions during medical emergencies. Through an online experiment with 35 providers, we compared two human-AI interaction strategies: (1) AI information synthesis and (2) AI information and recommendations. We found evidence supporting our second hypothesis that presenting AI information and recommendations improves the accuracy of decisions about life-saving interventions as compared to no AI support. In contrast, we did not find a significant difference in decision-making accuracy between AI information synthesis and no AI support. Participants often made decisions as patient information was being presented and synthesized on the AI support display. As a result, recommendations usually appeared after initial decisions, with some participants describing how the recommendations prompted them to check their decisions. Providers had an overall preference for AI information and recommendations compared to the other types of AI support. However, some providers had negative perceptions of the recommendations, and were concerned that the recommendations could be incorrect or lead to biased decision making. Although we found that providing AI information and recommendations improved decisions about life-saving interventions, we also identified two socio-technical barriers to implementing AI recommendations for time-critical decision making. First, decision-support systems used in emergency medicine will likely face a time-accuracy trade-off in presenting recommendations because patient information is dynamically obtained and providers often make decisions rapidly with limited information. Second, AI recommendations could potentially have negative effects on provider collaboration and agency. Based on the study findings, we proposed three implications for developing AI-enabled decision support for time-critical decision making: (1) support human decision makers in critically evaluating AI-generated information synthesis, (2) consider information-driven strategies, beyond only recommendations, to support providers in managing accuracy-time trade-offs, and (3) develop clear policies around the responsibility and liability of human decision makers when adopting AI-enabled tools. Our study contributes to the limited amount of empirical knowledge on the effectiveness of different human-AI interaction strategies in supporting domain experts during time-critical decision making.

\begin{acks}
We thank the physicians who participated in this research. We also thank Vidhi Shah for assisting in the design of the decision-support displays and Dr. Karen O'Connell for supporting the development of the vignettes. This research has been supported in part by the National Library of Medicine of the National Institutes of Health under grant number 2R01LM011834-05. This research was also supported in part by the National Science Foundation under grant number IIS-2107391 and the Graduate Research Fellowship Program under grant number 2041772. Any opinions, findings, conclusions or recommendations expressed in this material are those of the authors and do not necessarily reflect the views of the National Science Foundation or the National Institutes of Health.
\end{acks}

\bibliographystyle{ACM-Reference-Format}
\bibliography{references}

\appendix

\section{Participant Demographics}\label{appendix_demopgrahics}

\begin{table}[H]
  \caption{Demographic information for the design research participants (phase I).}
  \label{tab:interview_participants_full}
  \begin{tabular}{l|l|l|l|c}
    \toprule
    \textbf{Design Activity}&\textbf{ID}&\textbf{Specialty}&\textbf{Role}&\textbf{Years of Experience}\\
    \midrule
    Survey&1&Emergency Medicine&Attending&8\\
    \midrule
    Survey&2&Emergency Medicine&Attending&17\\
    \midrule
    Survey&3&Emergency Medicine&Fellow&2\\
    \midrule
    Survey&4&Emergency Medicine&Attending&10\\
    \midrule
    Survey&5&Emergency Medicine&Attending&6\\
    \midrule
    Survey&6&Emergency Medicine&Attending&10\\
    \midrule
    Survey&7&Emergency Medicine&Attending&23\\
    \midrule
    Survey&8&Emergency Medicine&Attending&8.5\\
    \midrule
    Survey&9&Emergency Medicine&Attending&5\\
    \midrule
    Survey&10&Emergency Medicine&Attending&30\\
    \midrule
    Survey&11&Emergency Medicine&Attending&20\\
    \midrule
    Survey&12&Surgery&Attending&15\\
    \midrule
    Survey&13&Emergency Medicine&Attending&7\\
    \midrule
    Survey&14&Emergency Medicine&Attending&33\\
    \midrule
    Survey&15&Emergency Medicine&Attending&18\\
    \midrule
    Survey&16&Emergency Medicine&Attending&14\\
    \midrule
    Survey&17&Emergency Medicine&Attending&30\\
    \midrule
    Survey&18&Emergency Medicine&Attending&13\\
    \midrule
    Survey&19&Surgery&Attending&5\\
    \midrule
    Interview&1&Emergency Medicine&Attending&30\\
    \midrule
    Interview&2&Emergency Medicine&Attending&18\\
    \midrule
    Interview&3&Emergency Medicine&Attending&14\\
    \midrule
    Interview&4&Emergency Medicine&Attending&6\\
    \midrule
    Interview&5&Emergency Medicine&Fellow&3\\
    \midrule
    Interview&6&Emergency Medicine&Attending&35\\
    \midrule
    Interview&7&Emergency Medicine&Attending&8\\
    \midrule
    Interview&8&Emergency Medicine&Attending&10\\
    \midrule
    Interview&9&Emergency Medicine&Attending&8\\
  \bottomrule
  \end{tabular}
\end{table}

  \begin{tabularx}{\textwidth}{l|l|l|c}
    \caption{Demographic information for evaluation participants (phase II).}\\
    \toprule
    \textbf{ID}&\textbf{Specialty}&\textbf{Role}&\textbf{Years of Experience}\\
    \midrule
    1&Emergency Medicine&Fellow&5\\
    \midrule
    2&Surgery&Nurse Practitioner&7\\
    \midrule
    3&Emergency Medicine&Attending&39\\
    \midrule
    4&Surgery&Resident&1.5\\
    \midrule
    5&Emergency Medicine&Attending&14\\
    \midrule
    6&Emergency Medicine&Fellow&3.5\\
    \midrule
    7&Emergency Medicine&Fellow&1\\
    \midrule
    8&Emergency Medicine&Fellow&4\\
    \midrule
    9&Emergency Medicine&Attending&8\\
    \midrule
    10&Emergency Medicine&Attending&11\\
    \midrule
    11&Emergency Medicine&Attending&3\\
    \midrule
    12&Surgery&Resident&7\\
    \midrule
    13&Surgery&Resident&6\\
    \midrule
    14&Surgery&Resident&5\\
    \midrule
    15&Surgery&Resident&4\\
    \midrule
    16&Surgery&Resident&7\\
    \midrule
    17&Surgery&Resident&4\\
    \midrule
    18&Surgery&Attending&22\\
    \midrule
    19&Surgery&Attending&30\\
    \midrule
    20&Surgery&Attending&25\\
    \midrule
    21&Critical Care&Attending&30\\
    \midrule
    22&Critical Care&Fellow&7\\
    \midrule
    23&Critical Care&Attending&14\\
    \midrule
    24&Emergency Medicine&Attending&27\\
    \midrule
    25&Surgery&Resident&5\\
    \midrule
    26&Critical Care&Attending&13\\
    \midrule
    27&Critical Care&Fellow&5\\
    \midrule
    28&Surgery&Resident&2\\
    \midrule
    29&Surgery&Resident&4\\
    \midrule
    30&Surgery&Resident&4\\
    \midrule
    31&Surgery&Resident&3\\
    \midrule
    32&Surgery&Resident&4\\
    \midrule
    33&Surgery&Resident&4\\
    \midrule
    34&Surgery&Resident&1\\
    \midrule
    35&Surgery&Resident&1\\
  \bottomrule
  \end{tabularx}

\section{Experiment Debriefing Questions} \label{appendix}
\begin{itemize}
    \item Please rank the types of AI support in order of effectiveness in supporting decision-making.
    \item Please rank the types of AI support in order of efficiency in supporting decision making.
    \item Please rank the types of AI support in order of overall preference.
    \item How did you use the information provided by the AI-support system (if at all)?
    \item How did you use the recommendation from the AI-support system (if at all)? 
    \item From 0\% to 100\%, how accurate did you find the AI recommendations to be?
    \item (Optional) What changes would you make to the design of the system (if any)?
    \item (Optional) What concerns do you have about receiving information from an AI system during medical events (if any)?
    \item (Optional) What concerns do you have about receiving recommendations from an AI system during medical events (if any)?
    \item (Optional) Did you face any technical difficulties during the experiment?
\end{itemize}

\section{Regression Tables} \label{appendix_regression_model}
\begin{table}[H]
  \caption{Diagnostic Accuracy (n=816) - Logistic Regression}
  \label{tab:diagnostic_accuracy}
  \begin{tabular}{l|l|l}
    \toprule
    \textbf{}&\textbf{Chisq}&\textbf{p-value}\\
    \midrule
    Type of AI Support&9.06&0.01*\\
    \midrule
    Specialty&7.44&0.02*\\
    \midrule
    "Borderline" Risk Prediction Score&69.52&<0.001***\\
    \midrule
    Training&2.03&0.15\\
    \midrule
    Order of Experiment Conditions&0.70&0.87\\
    \midrule
    Intervention Type&8.12&<0.01**\\
  \bottomrule
\end{tabular}
\end{table}

\begin{table}[H]
  \caption{Time to Initial Decision (n=816) - Linear Regression}
  \label{tab:diagnostic_accuracy}
  \begin{tabular}{l|l|l}
    \toprule
    \textbf{}&\textbf{F-statistic}&\textbf{p-value}\\
    \midrule
    Type of AI Support&0.14&0.87\\
    \midrule
    Specialty&0.74&0.54\\
    \midrule
    "Borderline" Risk Prediction Score&21.47&<0.001***\\
    \midrule
    Training&2.28&0.14\\
    \midrule
    Order of Experiment Conditions&2.68&0.06\\
    \midrule
    Intervention Type&0.19&0.66\\
  \bottomrule
\end{tabular}
\end{table}

\begin{table}[H]
  \caption{Time to Final Decision (n=816) - Linear Regression}
  \label{tab:diagnostic_accuracy}
  \begin{tabular}{l|l|l}
    \toprule
    \textbf{}&\textbf{F-statistic}&\textbf{p-value}\\
    \midrule
    Type of AI Support&1.07&0.34\\
    \midrule
    Specialty&0.77&0.52\\
    \midrule
    "Borderline" Risk Prediction Score&12.07&<0.001***\\
    \midrule
    Training&2.19&0.15\\
    \midrule
    Order of Experiment Conditions&2.70&0.07\\
    \midrule
    Intervention Type&1.00&0.32\\
  \bottomrule
\end{tabular}
\end{table}

\end{document}